\documentclass[prb,twocolumn,showpacs,amsmath,amssymb]{revtex4}

\usepackage{graphicx}
\usepackage{dcolumn}
\usepackage{bm}

\begin{document}


\title{A Systematic Study of the Superconducting and Normal State Properties of Neutron Irradiated MgB$_2$}

\date{\today}

\author{R. H. T. Wilke, S. L. Bud'ko, and P. C. Canfield}
\affiliation{Ames Laboratory US DOE and Department of Physics and Astronomy, Iowa State University, Ames, IA
50011}

\author{J. Farmer}
\affiliation{Missouri University Research Reactor, University of Missouri - Columbia, Research Park, Columbia, MO
65211}

\author{S. T. Hannahs}
\affiliation{National High Magnetic Field Laboratory, Florida State University, 1800 E. Paul Dirac Drive,
Tallahassee, Florida 32310}

\date{\today}

\begin{abstract}
We have performed a systematic study of the evolution of the superconducting and normal state properties of
neutron irradiated MgB$_2$ wire segments as a function of fluence and post exposure annealing temperature and
time. All fluences used suppressed the transition temperature, T$_c$, below 5 K and expanded the unit cell. For
each annealing temperature T$_c$ recovers with annealing time and the upper critical field, H$_{c2}$(T=0),
approximately scales with T$_c$. By judicious choice of fluence, annealing temperature and time, the T$_c$ of
damaged MgB$_2$ can be tuned to virtually any value between 5 and 39 K. For higher annealing temperatures and
longer annealing times the recovery of T$_c$ tends to coincide with a decrease in the normal state resistivity and
a systematic recovery of the lattice parameters.

\end{abstract}


\maketitle

\section{Introduction}

It has been well established that superconductivity near 40 K \cite{1} in MgB$_2$ is phonon mediated \cite{2,3}
with two distinct energy gaps \cite{4} [see reference 5 and works therein for a review of the basic properties of
MgB$_2$]. Much recent work has focused on understanding how structural defects and chemical substitutions affect
the superconducting properties. The goal is two fold; first, to understand the basic physics in this unique system
and second, to use this understanding as a basis for developing practical superconducting wires, tapes, or
devices.

Of the many elements investigated, only Al and C have shown concrete evidence of entering the structure \cite{6}.
Both electron dope the system and it is the the Fermi surface changes, rather than interband scattering, which are
believed to cause the suppression of T$_c$ \cite{3,7}. When aluminum substitutes for magnesium, H$^{\parallel
ab}_{c2}$ decreases while H$^{\bot ab} _{c2}$ remains constant or slightly increases, resulting in an decrease in
the anisotropy ratio, $\gamma$$_H$=H$^{\parallel ab}_{c2}$$/$H$^{\bot ab} _{c2}$. \cite{7,8,9}. These changes can
be understood as the result of changes in the Fermi surface topology of the $\sigma$ band \cite{10}. In the case
of carbon substitution for boron, the upper critical field is enhanced in both directions \cite{11,12}, with
H$^{\bot ab} _{c2}$ increasing more rapidly than H$^{\parallel ab}_{c2}$ leading to a decrease in the anisotropy
ratio \cite{7}. This enhancement of the upper critical field can not be explained in terms of changes in the Fermi
surface and is believed to result from an increase the scattering in the $\pi$ band \cite{7} in accordance with
the theory proposed by Gurevich \cite{13}.

Damaging or disordering of the sample using protons, heavy ions, neutrons, etc., is another route to
systematically changing the system. Of these possible routes, neutron irradiation offers the best avenue for
uniformly damaging bulk MgB$_2$. There are two main sources of damage from neutron irradiation of MgB$_2$. First,
fast neutrons deposit energy through inelastic collisions with atoms, creating thermal and dislocation spikes
\cite{24}. Second, $^{10}$B has a large capture cross section for lower energy neutrons  and readily absorbs these
thermal neutrons, subsequently $\alpha$ decaying to $^7$Li. Early neutron damage studies focused on irradiation of
powders or pressed pellets of MgB$_2$ containing natural boron \cite{14,15,16,17}.

The absorption of slow neutrons by $^{10}$B can lead to self shielding and prevents uniform damage in bulk MgB$_2$
samples significantly larger than the penetration depth. Under the assumption of linear absorption, the intensity
of the incident beam decreases exponentially within the sample, and is given by \cite{99}:

\begin{equation}
I(t)=I_o exp(-tsN)
\end{equation}

\noindent where I$_0$ is the initial intensity, t is the sample depth, s is the absorption cross section, and N is
the number density of atoms. Here, N corresponds to the number of $^{10}$B atoms per cm$^3$. A calculation of the
half depth in MgB$_2$ synthesized from natural boron, which consists of 19.9\% $^{10}$B, yields a depth of
approximately 130 $\mu$m. Samples with dimensions significantly larger than 130 $\mu$m will contain substantial
gradients of damage associated with slow neutron absorption. Therefore several different approaches have been
employed to minimize neutron capture, ensuring homogeneous damage throughout bulk samples. Thermal neutrons have
been blocked by a cadmium shield \cite{15,18,19} or the natural boron has been replaced with isotopically enriched
$^{11}$B \cite{20,21}.

An alternate approach is to damage MgB$_2$ objects with characteristic dimensions comparable to or smaller than
130 $\mu$m. In this case natural boron can be used and relatively homogeneous absorption of slow neutrons can be
assumed. In this paper we report on the isotropic irradiation of fully dense, 140 $\mu$m diameter, MgB$_2$ wires
with both thermal and fast neutrons. Although these fibers contain natural boron, they were exposed to an
isotropic fluence and had sample dimensions comparable to the penetration length, implying the defect structure
should be fairly uniform throughout the samples.

\section{Experimental Methods}

We synthesized fully dense, 140 micrometer diameter MgB$_2$ wire by reacting boron filaments with magnesium vapor
at 950 $^o$C for 36 hours. This technique is described in full detail elsewhere \cite{22}. Once synthesized, three
fibers, each approximately 1-2 cm in length, were sealed under a partial helium atmosphere in quartz ampoules with
dimensions of 3 mm I.D., 4 mm O.D., and approximately 3 cm long. A partial atmosphere of inert gas was necessary
to provide thermal contact with the cooling water, so as to prevent the filaments from overheating during the
irradiation. Helium was used due to its low neutron capture cross section. A 2.5 cm diameter, 4.7 cm long water
flooded aluminum can containing 25 ampoules was then exposed to an isotropic flux of reactor neutrons, consisting
of 98\% thermal neutrons (E=25.3 meV) and 2\% epithermal neutrons (ranging in energy up to 10 keV), at the
Missouri University Research Reactor (MURR) for time periods of 24, 48, 72, and 96 hours. The exposures
corresponded to fluences of $4.75 \times 10^{18}$, $9.50 \times 10^{18}$, $1.43 \times 10^{19}$, and $1.90 \times
10^{19}$ cm$^{-2}$ respectively (Table 1). Table 1 also includes estimates of the atomic percentage of boron
converted to lithium. The fluences were estimated using a cobalt doped aluminum flux wire. The percent of lithium
produced was estimated, under the assumption of a uniform fluence throughout the sample, by the product of the
density of $^{10}$B, the $^{10}$B neutron capture cross section, and the neutron fluence \cite{99}. Post exposure
anneals were performed with the samples still sealed within the quartz ampoules for temperatures up to 500 $^o$C
in a Lindberg model 55035 Mini-Mite tube furnace. In each case, we annealed an ampoule at a given temperature for
a set period of time. Upon removal from the furnace, the ampoules were quenched in air and then opened in order to
perform measurements on the individual wires. Each data point (i.e. each fluence/annealing time/annealing
temperature) is from a single sample that was annealed only once. In some cases we measured several samples which
underwent nominally the same conditions and this was done either by using different wires within a given ampoule
or from annealing additional ampoules under the same profile. In all cases measurements were performed on samples
which underwent only a single anneal.

The boron filaments used in this study contain a tungsten boride core. Fast neutrons colliding with $^{182}$W
atoms, which have a natural abundance of 26.3\%, can be absorbed into the nucleus causing the emission of a proton
and transforming the tungsten into $^{182}$Ta. $^{182}$Ta $\beta$ decays back to $^{182}$W, with a half life of
181 days. As a result the filaments were mildly radioactive and required appropriate safety measures in handling.

Powder x-ray diffraction (XRD) measurements were made at room temperature using CuK$\alpha$ radiation in a Rigaku
Miniflex Diffractometer. A silicon standard was used to calibrate each pattern. The experimentally determined Si
peak positions were found to be offset from their known values by a constant amount. Within each spectra, the
peaks varied about some constant offset and this variation was used to estimate experimental uncertainty in the
lattice parameters. Lattice parameters were determined from the position of the (002) and (110) peaks. DC
magnetization measurements and magnetization hysteresis loops were performed in a Quantum Design MPMS-5 SQUID
magnetometer. Transport measurements were done using a standard AC four probe technique, with platinum wires
attached to the samples with Epotek H20E silver epoxy. In a previous study \cite{187} we note that during the
synthesis of the boron filament, the tungsten reacts with B in the gas stream forming a host of tungsten boride
compounds. As a result the core is not a low resistance W wire, but a much more highly resistive alloy of the
refractory boride materials W$_x$B$_y$ that, due to expansion associated with the formation of the MgB$_2$ is
segmented. In addition to the transmutation of some of the W to Ta, the presence of natural boron in the core will
lead the presence of defects caused by the transmustion of $^{10}$B to $^7$Li just as in the MgB$_2$ wire.
Additionally, the cross section of the core is only 2\% of the cross section of the wire. We therefore take the
measurement of the resistivity to represent the resistivity of the MgB$_2$. (This assumption is valid as long as
the resistivity of the core remains above $\sim 0.5 \mu\Omega$~cm.  Given the non-stoichiometry, radiation damage,
transmutation and segmentation of the core we feel this is a safe assumption.) Resistivity versus temperature
measurements in applied magnetic fields up to 14 T were carried out in a Quantum Design PPMS-14 system and
resistivity versus field was measured up to 32.5 T in a DC resistive magnet using a lock-in amplifier technique at
the National High Magnetic Field Laboratory in Tallahassee, Florida.

\section{Fluence and Annealing Studies}

Normal state and superconducting properties were found to be a function of fluence, annealing time, and annealing
temperature. We present the data in the following order: (a) structural studies on as-damaged wires, (b) 24 hour
anneals at varying temperatures for a fluence of $4.75 \times 10^{18}$ cm$^{-2}$, (c) Variable time anneals for a
fluence of $4.75 \times 10^{18}$ cm$^{-2}$, (d) 24 hour anneals for all measured fluence levels, and (e) 1000 hour
anneals on all four fluence levels. With these cuts through the multi-parameter, fluence, annealing temperature,
annealing time phase space we can start to describe the effects of neutron damage on MgB$_2$ samples made with
natural boron.

\subsection{Structural Studies of As Damaged Samples}

The x-ray (002) and (110) peaks for an un-damaged wire and the entire set of as-damaged samples are plotted in
figure 1a. The lowest exposure level, $4.75 \times 10^{18}$ cm$^{-2}$, shows an anisotropic expansion of the unit
cell. The a- lattice parameter increases from 3.0876(5) $\AA$ in the un-damaged sample to 3.0989(2) $\AA$, an
increase of 0.0113(7) $\AA$ or 0.37\%. The c- lattice parameter increases from 3.5209(7) $\AA$ to 3.5747(2) $\AA$,
an increase of 0.0538(9) $\AA$ or 1.02\%. Similar anisotropic expansion of the unit cell was seen by Karkin et.
al. \cite{14}. The authors report lattice parameter increases of $\Delta$a=0.0075 $\AA$ or 0.24\% and
$\Delta$c=0.0317 $\AA$ or 0.9\% for a fluence of $1 \times 10^{19}$ cm$^{-2}$ thermal neutrons and $5 \times
10^{18}$ cm$^{-2}$ fast neutrons. For irradiation of isotopically enriched Mg$^{11}$B$_2$, little change was seen
in the a- lattice parameter up to a fluence level of 10$^{17}$ cm$^{-2}$ \cite{20}. For this fluence level the
authors report a 0.008 $\AA$ or 0.23\% increase in the c- lattice parameter relative to an undamaged sample. Due
to the large differences in $\Delta$a and $\Delta$c reported for different irradiation conditions, we performed
systematic study of the lattice parameters as a function of fluence level, annealing temperature, and annealing
time for samples all prepared in the same manner.

Relative to the $4.75 \times 10^{18}$ cm$^{-2}$ fluence level, filaments exposed to a fluence of $9.50 \times
10^{18}$ cm$^{-2}$ show a further expansion of the unit cell, with the a- and c-lattice parameters reaching
3.1017(7) $\AA$ and 3.5805(10) $\AA$, respectively. The response to further increases in exposure qualitatively
changes. For an exposure level of $1.43 \times 10^{19}$ cm$^{-2}$ the (002) peak broadens substantially and the a-
and c- lattice parameters contract relative to the samples exposed to the $9.50 \times 10^{18}$ cm$^{-2}$ fluence
level. Increasing the fluence further to $1.90 \times 10^{19}$ cm$^{-2}$ results in a further contraction of both
the a- and c- lattice parameters with the (002)peak remaining broad. The x-ray scan for the as-damaged, $1.90
\times 10^{19}$ cm$^{-2}$ fluence level sample, from 2$\theta$=20-65$^o$ is plotted in figure 1b. The peak widths
for the MgB$_2$ (hk0) peaks are within a factor of two of the widths for neighboring Si peaks, whereas the MgB$_2$
(001) peak is not resolvable and the (002) peak is more than five times as broad as the Si peaks. It should be
noted that in pure, undamaged MgB$_2$ all x-ray peaks have full width at half maximum (FWHM) values comparable to
those of the Si standard. For example, the peaks used to determine the lattice parameters in the undamaged wires
used in this study, the (002) and (110), had FWHM values of 0.216$^o$ and 0.221$^o$ respectively. The neighboring
Si peak, the (311) peak which occurs at 56.110$^o$ 2$\theta$, had a FWHM of 0.190$^o$.

In the case of a fluence level of $1.90 \times 10^{19}$ cm$^{-2}$, the MgB$_2$ (100) peak, occurring at
2$\theta$=33.397$^o$ has a full width at half maximum (FWHM) of 0.216$^o$, in comparison to the FWHM of 0.180$^o$
for the Si (111) peak at 2$\theta$=28.422$^o$. Similarly the MgB$_2$ (110) and Si (311), which occur at 59.686$^o$
and 56.110$^o$ 2$\theta$, had FWHM values of 0.291$^o$ and 0.163$^o$ respectively. In contrast, the MgB$_2$ (002)
peak has a FWHM of 1.335$^o$. The resultant correlation lengths, given by $\Delta$2$\theta$ =
$\lambda/d~cos(\theta)$, are approximately 4000 $\AA$ in the plane and 700 $\AA$ along the c-direction. The mixed
(101) peak has a FWHM of 1.045$^o$ and a correlation length near 1000 $\AA$, values intermediate between the two
extrema. The two highest fluence levels have resulted in an anisotropic change in the correlation length, with a
decrease in long range order between the boron planes. Since the (002) peak has substantially broadened for higher
fluence levels, tracking the trend in $\Delta$c becomes harder due to the inherent decrease in accurately defining
peak position.

\subsection{24 Hour Anneals of Samples Exposed to a Fluence of
$4.75 \times 10^{18}$ cm$^{-2}$}

As-damaged samples showed a suppression of T$_c$ to below 5 K. This is a much larger suppression than seen upon
irradiating Mg$^{11}$B$_2$ \cite{20}: the authors report a T$_c$ of 12.2 K for a fluence of $3.9 \times 10^{19}$
cm$^{-2}$. Irradiation of MgB$_2$, i.e. with natural boron, using a fast neutron fluence of $2 \times 10^{18}$
cm$^{-2}$ gave a T$_c$ of 30.2 K \cite{16}. Thus damage induced via neutron capture and subsequent alpha decay
appears to play a significant role in suppressing superconductivity. To further investigate the effects of uniform
irradiation on samples containing natural boron we have performed a systematic study of the normal state and
superconducting properties as a function of annealing temperature for samples annealed for 24 hours.

We annealed a set of wires, exposed to a fluence of $4.75 \times 10^{18}$ cm$^{-2}$, for 24 hours at 100 $^o$C,
150 $^o$C, 200 $^o$C, 300 $^o$C, 400 $^o$C, and 500 $^o$C. X-ray measurements indicate the initial expansion of
the unit cell could be systematically reversed by subsequent annealing (Figure 2), with the $\Delta$a and
$\Delta$c values decreasing with increasing annealing temperature (Figure 2b). The a-lattice parameters is
completely restored after annealing at 400 $^o$C whereas the c- lattice parameter appears to be saturating at a
value near 0.6\% larger than that of the undamaged sample.

The superconducting transition temperatures were determined using a 1\% screening criteria in zero field cooled DC
magnetization measurements (Figure 3a) and an onset criteria in resistivity versus temperature measurements
(Figure 3b). The magnetic transitions for the entire set are fairly sharp, typically achieving 95\% of the
saturated value 2.5 K below the 1\% criteria used for T$_c$. This suggests a relatively homogeneous damaging of
the sample. We were unable to obtain reliable transport measurements on as damaged as well as the 100 $^o$C
annealed samples due to the 120 $^o$C temperature required to cure the silver epoxy contacts. The 30 minute 120
$^o$C cure of the epoxy, as well as the possible exotherm associated with the epoxy curing, placed the sample into
a poorly defined annealed state. As shown in figures 3a and 3b, superconductivity is restored by the annealing
process. $\Delta T_c$, defined as the difference between the undamaged $T_c$ and that of the annealed sample,
monotonically approaches zero as the annealing temperature is increased (Figure 4).

The zero field temperature dependent resistivity data are plotted in figure 3b. The sample annealed at 150$^o$C
has a low temperature normal state resistivity slightly above 6 $\mu\Omega~$cm, which is an order of magnitude
larger than the undamaged value of 0.2 $\mu\Omega~$cm. After annealing at a temperature of 200$^o$C, $\rho$
increases by another four fold. Further increases in the annealing temperature result in a monotonic decrease of
the normal state resistivity, to 1.6 $\mu\Omega~$cm at an annealing temperature of 500$^o$C. Although we were able
to perform a measurement on only a single sample, this increase in $\rho$ at 200$^o$C is thought to be a real
effect, as it also manifests itself in normalized $\rho$/$\rho$(300 K) plots, indicating it is not associated with
a pathologic geometric problem such as cracks. Additionally, un-damaged samples consistently showed T$_c$ values
near 39 K and normal state resistivities in the 0.2-0.4 $\mu\Omega~$cm range, values in agreement with previously
published results \cite{22,730}, suggesting the effect is the result of the irradiation and post-exposure
annealing process. Such an anomalous increase in resistivity as a function of annealing temperature was also
observed in the case of neutron irradiation on carbon doped MgB$_2$ wire segments \cite{991}. The initial
annealing may result in a change in the defect structure. For example, vacancies don't necessarily immediately
recombine with interstitials to eliminate defects. Vacancies can initially cluster together to form higher order
defects, which can be relatively stable \cite{24}. Thus the initial increase in resistivity may result from a
reorganization of defects into higher order clusters which may enhance scattering.

Insight into the nature of the defects can be obtained by tracking changes in the parameter
$\Delta$$\rho$=$\rho$(300 K)-$\rho$(40 K). Figure 3d plots $\Delta$$\rho$ as a function of the transition
temperature. $\Delta$$\rho$ decreases monotonically as a function of the transition temperature. This result is in
contrast to the results of He ion irradiation on MgB$_2$ thin films \cite{476}, but in agreement with the
predictions of Mazin et al. \cite{1215}, who claimed that as a material is driven into the dirty limit, the
conductivity in the pi band decreases, causing normal state electric transport to switch from the pi to the sigma
band. Indeed, for the extreme clean and extreme dirty limits they obtain values ranging from near 10 to near 35
$\mu\Omega~$cm respectively, consistent with our experimental observations of 6 and 35 $\mu\Omega~$cm.

An alternate framework for understanding this monotonic change in $\Delta \rho$  is to recall that the sample is
polygrain with random orientation of the crystallites.  Given that the electrical resistivity of MgB$_2$ is
anisotropic to start with, combined with the fact that damage appears to more aggressively affect the $c$-axis
periodicity, it is possible that for lower $T_c$ values an increasing number of grain orientations will present
high resitivity paths to the current.  This will effectively reduce the cross sectional area of the wire sample.
This would result in a similar behavior to what has been found for samples that have either poor connectivity or
gross impurities on the grain boundaries \cite{1216}.

The upper critical field was determined using an onset criteria in resistivity versus temperature in applied
fields up to 14 T (a representative set is shown in figure 5a) and, in the case of the 500$^o$C anneal, resistance
versus field sweeps up to 32.5 T (Figure 5b). The upper critical field curves nest, forming a sort of Russian doll
pattern, with H$_{c2}$(T=0) approximately scaling with T$_c$ (Figure 6). The curves for samples annealed at
temperatures up to 200$^o$C do not show any positive curvature near T$_c$ and are qualitatively similar to single
gap superconductors with Werthamer, Helfand, and Hohenberg (WHH) \cite{25} like behavior. Experimentally
determined H$_{c2}$(T=0) values for the 150$^o$C, 200$^o$C, and 300$^o$C anneals are 2.9 T, 4.7 T, and 7.3 T
respectively. Using the formula H$_{c2}$(T=0)=0.69T$_c$dH$_{c2}$/dT we obtain estimates of 2.9 T, 4.3 T and 5.9 T.
Thus whereas only in the cases of the 150$^o$C and 200$^o$C anneals we can fit H$_{c2}$(T) with WHH behavior, the
deviations increase with the annealing temperature, suggesting that the bands may become fully mixed only when
T$_c$ is suppressed to near 10 K. Single gap behavior has been inferred from specific heat measurements on
irradiated samples containing isotopically enriched $^{11}$B which had T$_c$ near 11 K \cite{1218}. The 300$^o$C,
400$^o$C, and 500$^o$C anneals exhibit positive curvature near T$_c$ that is similar to what is found in pure
MgB$_2$. The 500$^o$C anneal data show that either the undamaged H$_{c2}(T)$ is restored or that there a slight
increase in H$_{c2}$(T=0), rising from approximately 16 T in the undamaged case to near 18 T.

This behavior in H$_{c2}$ differs from other types of neutron damaging studies. For the case of irradiating
isotopically enriched Mg$^{11}$B$_2$. Putti et. al found a fairly substantial increase in H$_{c2}$ values when
T$_c$ was in the 36-38 K range \cite{18}. For a sample with T$_c$ = 36.1 K, they report H$_{c2}$(T=12K) of 20.3 T.
For unshielded irradiation of MgB$_2$ containing natural boron, Eisterer et. al reported that, relative to an
undamaged sample with a T$_c$ just below 38 K, a sample with a suppressed T$_c$ of approximately 36 K had a near
doubling of the slope dH$_{c2}$/dT in the linear regime above 2 T \cite{15}. In both of these cases the fluences
used were an order of magnitude lower than in our experiment and no post exposure annealing was performed.

Figure 7 presents the critical current densities for the series of 24 hour anneals at 5 K and 20 K as determined
by the Bean critical state model \cite{23} from magnetic hysteresis loops. At both 5 K and 20 K, J$_c$ is
suppressed for samples annealed at 300 $^o$C and below, presumably due to the larger reduced temperature (T/T$_c$)
of the measurements. In field J$_c$ values become enhanced at both 5 K and 20 K for the 400 $^o$C and 500 $^o$C
anneals, and show a slight fishtail or second peak behavior. This is believed to be the first time such behavior
has been observed in polycrystalline samples. The fish tail effect has been seen in fast neutron irradiation of
MgB$_2$ single crystals \cite{19}. The authors report a second peak in the magnetization hysteresis loops for
H$\bot$ab, which became more pronounced with increasing fluence up to $4V \times 10^{17}$ cm$^{-2}$. No second
peak was observed for H$||$ab. Low level thermal neutron fluence, of order 10$^{14}$-10$^{15}$ cm$^{-2}$, of
polycrystalline MgB$_2$ showed an enhancement in flux pinning, but no second peak or fishtail behavior \cite{16}.
The infield enhancement of J$_c$ for the sample annealed at 500$^o$C pushes the point at which J$_c$ crosses
10$^4$ A/cm$^2$ at 20 K out to near 1.5 T. This increase in field roughly doubles the crossing point of the
undamaged wire, but falls well below best reports in literature of 10$^4$ A/cm$^2$ at near 5 T reported for 10
weight percent addition of SiC to MgB$_2$ \cite{27}.

Other authors have analyzed critical current densities in neutron irradiated samples using a percolative model
that considers, as free parameters, the upper critical field, the absolute value of the critical current density,
the anisotropy of the upper critical field, and the percolation threshold \cite{1217,1219,1230}. In reference
\cite{1230} the authors extend the model to include different contributions from grain boundary and precipitate
pinning. The model yields extremely good fits to the data, from which the authors conclude that the anisotropy of
H$_{c2}$ heavily influences the field in which the material can carry large amounts of current \cite{1217,1219}
and defects introduced by the irradiation process are responsible for the enhancement of the pinning strength
\cite{1230}. It is worth noting that these reports do not show the significant change in J$_c$(0) that we observe
(Figure 7). Presumably two effects are causing the increase as a function of annealing temperature and hence
T$_c$. In using the percolation model, each of the authors kept the percolation threshold fixed in the 0.2-0.3
range. If we are observing large changes in the connectivity in our samples as a function of annealing
temperature, as inferred from the $\Delta$$\rho$ data, then the initial decrease in J$_c$(0), as observed in the
low temperature anneals, is the result of the formation of weak links which limit the overall critical current
density. As the annealing temperature is increased the connectivity between grains is improved, thereby returning
J$_c$(0) towards the undamaged value. In the case of the 500$^o$C anneal, J$_c$(0) exceeds that of the undamaged
sample, suggesting that the defects which remain following post-exposure annealing act as effective pinning sites.
If defects caused by irradiation are 0.1-10 nm in size, as reported in reference \cite{1217}, then following the
annealing process, many of the defects should be of order the coherence length \cite{69}, making them effective
flux pinning centers.

A plot of J$_c$ as a function of reduced field (H/H$_{c2}$) shows that for the higher temperature anneals, the
fraction of H$_{c2}$ in which samples can maintain in excess of 10$^3$ A/cm$^2$ is greatly enhanced (Figure 7c and
7d). Such enhancement is consistent with an increase in the strength of defect pinning \cite{1230}. That the low
temperature anneals all show J$_c$ dropping below 10$^3$ A/cm$^2$ at similar reduced field values suggests the
defects are ineffective in enhance critical current densities in highly disordered samples. This could be the
result of, for example, overlap of defects, making them to large to be effective flux pinning sites.

\subsection{Variable Time Anneals on Samples Exposed to a
Fluence of $4.75 \times 10^{18}$ cm$^{-2}$}

The annealing time for the set of $4.75 \times 10^{18}$ cm$^{-2}$ fluence samples was varied to further probe the
characteristics of damage induced by neutron irradiation. An extensive set of anneals was carried out at 300$^o$C,
consisting of 1/3, 1, 3, 6, 24, 96, and 1000 hours. For annealing temperatures of 200$^o$C and 400$^o$C,
measurements were done on 1, 24, and 96 hour anneals. Two anneals were performed at 500$^o$C, one for 24 hours and
the other for 1000 hours.

The magnetic transitions for the entire set of samples annealed at 300$^o$C are shown in figure 8. Annealing at
this temperature for only 0.33 hours raised T$_c$ from below 5 K to slightly above 19 K. Therefore, the defects
causing the suppression of superconductivity must have a fairly low activation energy. For defects which can be
annealed by a single activated process with a constant activation energy, the rate of change of the defect
concentration is given by \cite{24}:

\begin{equation}
dn/dt=-F(n)K_0 e^{-E_a / k_B T}
\end{equation}

\noindent where $n$ is the defect concentration, $F(n)$ is some continuous function of $n$, $K_0$ is a constant,
$E_a$ is the activation energy, $k_B$ is Boltzman's constant, and $T$ is the temperature. If we assume random
diffusion, then $F(n)=n$, and the defect concentration decreases exponentially with time:

\begin{equation}
n=n_0 e^{-C t}
\end{equation}

\noindent where $n_0$ is the initial defect concentration, $C=K_0~e^{E_a / k_B T}$ is a rate constant, and $t$ is
time.

While $\rho_0$ can be taken as one measure of the defect concentration, we have already shown non-monotonic
behavior in $\rho_0$ as a result of the annealing process. Another variable that can be used to defect
concentration is $T_c$, which systematically varies with both annealing temperature and annealing time. We
therefore take $\Delta T_c$ as a measure of the defect concentration, although it should be reiterated that the
nature of the defects may be changing as we increase the annealing temperature and time.  Such linearity is not
uncommon and can be related to Abrikosov-Gorkov-like theory of pair breaking in anisotropic superconductors.
\cite{igor} It should be noted, though, that we do this simply in an attempt to extract an approximate value of
the activation energy.

We see a exponential decay behavior in $\Delta T_c$ for samples annealed at 200$^o$C, 300$^o$C and 400$^o$C
(Figure 9). Since we only have two samples annealed at 500$^o$C, which are longer time anneals and thus presumably
would fall further out on an exponential tail, we can not extract a meaningful value from an exponential fit of
these data. For samples which do show an exponential behavior in the decrease of $\Delta T_c$ (and by presumption)
the defect density as a function of annealing time the activation energy, assuming a single activation process,
can be determined by the so-called cross-cut procedure \cite{24}. This involves comparing the annealing time for
which different temperature anneals reached the same defect density, i.e. $\Delta T_c$. If a $\Delta T_c$ is
reached by annealing at a temperature $T_1$ for a time $t_1$ and by annealing at a temperature $T_2$ for a time
$t_2$, then the activation energy is related to these quantities by:

\begin{equation}
ln \frac{t_1}{t_2} = \frac{E_a}{k_B} (\frac{1}{T_1} - \frac{1}{T_2}).
\end{equation}

Unfortunately, no overlap region in $T_c$ exists for our set of data. To achieve an overlap region in the
experimental data we'd need to shorten the annealing times below 0.33 hours or extend them to well beyond 1000
hours. Short time anneals, less than 0.33 hours, are not feasible as we'd be unable to ensure thermal equilibrium
of the samples within the furnace for such a short time. Extending the annealing an additional order of magnitude,
to 10$^4$ hours, is simply not practical. We can still estimate the activation energy by extrapolating the $\Delta
T_c$ curves at each temperature to shorter and longer times so as to create overlap regions (Figure 9).
Calculating $E_a$ from the overlap between the 200$^o$C and 300$^o$C curves and between the 300$^o$C and 400$^o$C
curves yields values of 1.90 eV and 2.15 eV respectively.

An alternative approach to determining the activation energy is to use the ratio of slopes method \cite{106}.
Experimentally a set of samples is annealed isothermally for different times at a temperature $T_1$ and an
identical set of samples is annealed at a different temperature $T_2$. The differing temperatures result in
different time evolution of $\Delta T_c$, and hence, different slopes, $d\Delta T_c/dt$. The ratio of the slopes
for the two different temperatures at the point where both annealing temperatures have yielded the same $\Delta
T_c$, is related to the activation energy by:

\begin{equation}
\frac{d\Delta T_{c1}}{dt_1}/\frac{d\Delta T_{c2}}{dt_2} = exp( \frac{E_a}{k_B} (\frac{1}{T_1}-\frac{1}{T_2}))
\end{equation}

\noindent where the subscripts 1 and 2 refer to temperatures $T_1$ and $T_2$. By comparing the slopes from the
200$^o$C and 300$^o$C anneals and those from the 300$^o$C and 400$^o$C anneals, we obtain estimates of 1.07 eV and
1.63 eV, respectively. It should be noted that, while these estimates were made using real data, rather than
extrapolations as in the previous calculation, there is inherent inaccuracy in such a calculation due to the lack
of a true overlap in $\Delta T_c$. Additionally, the low density of data points limits our ability to accurately
determine the linear slope from the $\Delta T_c$ versus time curves where $\Delta T_c$ tends to decay
exponentially.

Depending on the calculation method and data set used, we obtain activation energies ranging from 1.07 eV to 2.15
eV. It is likely that some of the variation is real, as, in these heavily damaged samples, there exist both point
defects and defect complexes. The annealing of point defects is expected to have a lower activation energy than
the dissolving of the defect complexes. While we can not assign definitive values for the activation energies of
these two processes, merely stating a single activation energy hides some of the rich complexity underlying the
annealing process in these heavily damaged samples. It should be noted that the activation energies in these
neutron irradiated MgB$_2$ samples are the same order of magnitude as those for annealing quenched in defects out
of gold \cite{106}. That our samples yielded a relatively small spread in activation energies and were comparable
in magnitude to values associated with the annealing of simple defects in other metals suggests the defects within
the neutron irradiated MgB$_2$ samples are being annealed by single activation processes.

For the set of wires annealed at 300$^o$C, critical current densities at $T=5$ K and low fields approximately
scale with annealing time and hence $T_c$ (Figure 10). The field at which $J_c$ drops below 10$^4$ A/cm$^2$
increases by approximately a factor of two when the annealing time is increased from 1 hour to 24 hours. Extending
the annealing time further to 1000 hours has little effect on low field $J_c$ values. For annealing times of 6
hours and shorter (for the sake of clarity, data for 1/3, 3, and 6 hours are not shown) $J_c$ was found to
monotonically decrease as a function of applied field. After a 24 hour anneal, $J_c$ begins to flatten above 3 T
at a value of approximately 500 A/cm$^2$. Extending the annealing time to 96 and 1000 hours results in the
emergence of a clear second peak at an applied field slightly above 3 T. The curves for the 96 and 1000 hour
anneal are virtually identical and reflect the small 1.5 K or 6\% increase in T$_c$ resulting from the order of
magnitude longer annealing time.

\subsection{Variable Fluence Levels, 24 Hour Anneal}

We annealed samples from each of the damaged levels for 24 hours at 200$^o$C, 300$^o$C, 400$^o$C, and 500$^o$C.
The (002) and (110) x-ray peaks for samples of all four damage levels annealed at 300$^o$C for 24 hours are given
in figure 11. Annealing at 300$^o$C for 24 hours did not restore long range order along the c-axis in the samples
exposed to the two highest fluence levels of $1.43 \times 10^{19}$ cm$^{-2}$ and $1.90 \times 10^{19}$ cm$^{-2}$.
For these 24 hour anneals, up to an annealing temperature of 400$^o$C the (002) peak maintains a FWHM above 1$^o$,
which corresponds to a structural coherence length of approximately 1000 $\AA$ (Figure 12). Using H$_{c2}$(T=0) to
determine the supercoducting coherence length, one obtains for pure MgB$_2$ a superconducting coherence length
near 50 $\AA$ \cite{69}. Measurements on the coherence length within the $\pi$ band only yield a larger value of
approximately 500 $\AA$ \cite{823}. While there is shorter range structural order along the c- direction in these
heavily irradiated samples, the structural coherence length is still at least a factor of two larger than the
superconducting coherence length, regardless of the estimate of the superconducting coherence length used. It is
therefore not unexpected that superconductivity exists in samples with degraded long range order. Only after the
temperature reaches 500$^o$C does the correlation length along the c- direction exceed 1000 $\AA$ (Figure 12). The
relative shift of the a- and c- lattice parameters for the various heat treatments is plotted in figure 13.

Magnetization measurements were performed on all of the annealed samples (Figure 14a). It should be noted that for
the fluences greater than $4.75 \times 10^{19}$ cm$^{-2}$, as damaged samples showed no signs of superconductivity
down to 2 K. A summary of the transition temperatures for each of the annealed samples, given in terms of $\Delta
T_c$, is plotted in figure 14b. In general terms, the higher fluence levels lead to lower superconducting
transition temperatures for a given temperature post-exposure anneal. For a given fluence level, higher annealing
temperatures yield higher T$_c$ values. That is, the samples exposed to higher fluence levels behave in a
qualitatively similar manner to lowest level, but the added exposure leads to increased defect densities which
manifest themselves in terms of lower transition temperatures for a given annealing profile.

\subsection{Long Time Annealing Studies}

The conversion of $^{10}$B to $^{7}$Li through neutron absorption and subsequent alpha decay introduces the
possibility of observing the effects of lithium doping MgB$_2$. In order to distinguish the effects of Li doping
from those associated with structural defects introduced through inelastic collisions between neutrons and the
underlying lattice it is necessary to minimize the density of these defects. As shown, by annealing for long times
at high temperatures the number of structural defects can be systematically reduced. Since Li can't be annealed
away, the resultant superconducting and normal state properties should increasingly reflect the effects of Li
doping. The density of Li atoms produced through the transmutation of boron can be estimated from the formula
\cite{99}:

\begin{equation}
n_{Li} = n_{B} s f
\end{equation}

\noindent where $n_{Li}$ is the density of Li atoms, $n_{B}$ is the density of B atoms, $s$ is the absorption
cross section, and $f$ is the fluence level. Computing the corresponding atomic percentages yields an increase
from 0.37\% to 1.48\% as the fluence in increased from $4.75 \times 10^{18}$ cm$^{-2}$ to $1.90 \times 10^{19}$
cm$^{-2}$ (Table 1).

We therefore annealed all four of the damage levels for 1000 hours at 500$^o$C. Normalized magnetization curves
for this series are plotted in figure 15a. In each case the transition temperatures have increased relative to
samples annealed at 500 $^o$C for 24 hours as can be seen by plotting $\Delta T_c$ versus time (Figure 15b). Using
a linear fit on the semi-log plot, the resultant slopes are much lower than was seen for the $4.75 \times 10^{18}$
cm$^{-2}$ fluence level samples annealed for various times at 200$^o$C, 300$^o$C, and 400$^o$C (Figure 9),
indicating $\Delta T_c$ is beginning to saturate above 24 hours. Since we don't have any intermediate time points,
we can't determine if we've achieved fully saturated $\Delta T_c$ values at 1000 hours, but we can take the 1000
hour anneals at 500$^o$C as an upper limit on the effects of lithium doping.

Transport measurements were performed in order to determine the normal state resistivity and temperature
dependance of $H_{c2}$ for these samples. We were unable to contact the samples exposed to the two highest dose
levels. The normal state resistivity (T=40 K) for the sample exposed to a fluence of $4.75 \times 10^{18}$
cm$^{-2}$ and annealed for 1000 hours was approximately 5.3 $\mu\Omega~$cm, which is more than three times the 1.6
$\mu\Omega~$cm measured on the sample annealed for 24 hours. The 1000 hour anneal sample also had a lower residual
resistivity ratio (RRR), 3.3 versus 5.9, indicating the increase in $\rho_0$ isn't merely an artifact associated
with possible geometric effects (cracks within the sample). The $9.50 \times 10^{18}$ cm$^{-2}$ fluence level
sample, which was annealed for 1000 hours, had a normal state resistivity of 6.6 $\mu\Omega~$cm and RRR=5.0.
Karkin et al. saw increases in $\rho$ for annealing temperatures above 300$^o$C, which they attributed to changes
in inter-grain transport \cite{14}. It is possible that by annealing at an elevated temperature for 1000 hours we
have degraded the inter-grain connectivity in some fashion, leading to the observed increases in resistivity.

The temperature dependance of H$_{c2}$ is plotted in figure 16. It was found that for the $4.75 \times 10^{18}$
cm$^{-2}$ fluence level $H_{c2}$(T) data for a 1000 hour anneal sample was comparable to that of the 24 hour
anneal sample, both showing possible, slight enhancement relative to the pure sample. In the case of the 1000 hour
anneal on the $9.50 \times 10^{18}$ cm$^{-2}$ fluence level, H$_{c2}$ had some spread in the data, but
extrapolated to 18-19 T at zero Kelvin. The upper critical fields of the two lowest fluence levels annealed at
500$^o$C for 1000 hours appear to be similar, with slight differences arising due to inherent sample to sample
variation.

\section{Discussion}

The initial irradiation of MgB$_2$ wire segments results in an increase in the size of the unit cell and
suppression of the superconducting transition temperature. Post exposure annealing tends to return both the
lattice parameters and T$_c$ towards their undamaged values. If the superconducting properties of the neutron
irradiated samples were purely a result of changes in the unit cell dimensions, correlations should exist between
the superconducting properties of neutron irradiated MgB$_2$ and pure MgB$_2$ placed under external pressure.
Since an expansion is qualitatively analogous to an effective negative pressure one would anticipate the the
changes in T$_c$ to be an extension to negative pressure of the results attained for the application of positive
pressure. Application of external pressure has been shown to compress the unit cell and suppress T$_c$ \cite{34}.
Thin films grown epitaxially on (0001) sapphire substrates exhibited an enhanced T$_c$ above 41 K which was
attributed to tensile strain \cite{35}. In the case of neutron irradiated MgB$_2$, the expansion of the lattice
parameters coincides with a decrease in T$_c$ and the evolution of T$_c$ as a function of $\Delta$a, $\Delta$c,
and V/V$_0$ behaves differently than MgB$_2$ under pressure (Figure 17). In the case of externally applied
pressure, it is believed that the changes in the frequency of the E$_{2g}$ phonon mode are responsible for the
suppression of T$_c$ \cite{34,36}. With such dramatically different behavior between the neutron irradiation and
pressure results, it is clear that the changes in T$_c$ can not simply be linked to changes in the unit cell
volume.

In addition to the structural changes, neutron irradiation also introduces a chemical impurity into the system
through the $^{10}$B neutron capture and subsequent alpha decay to $^7$Li. The amount of lithium produced in this
manner is, however, quite small and for all but the 1000 hour anneals at 500$^o$C, we can not sort the possible
effects of such from the effects of the structural perturbations. The atomic percentage of B converted to Li
through nuclear processes for all four exposure levels is estimated to be on the order of 1\% (Table 1).
Therefore, for lower temperature and/or shorter time anneals the effects of Li production are presumably
insignificant next to the changes resulting from structural damage caused by inelastic collisions between the fast
neutrons, emitted alpha particles, and recoiled $^7$Li atoms  and the underlying lattice. It is worth noting,
though, that whereas low level Li substitution for Mg was proposed as a possible route to increasing T$_c$
\cite{84}, our data show suppressed T$_c$ values for our 1000 hour annealed samples. This does not preclude the
possibility that Li substitution could, under other circumstances, raise T$_c$, but it certainly does not support
it. It should be noted though that this form of lithium doping is far from the ideal one. In the case of
transmutation of $^{10}$B the resulting sample is MgLi$_x$B$_{2-x}$ with boron vacancies and no clear site for the
lithium. The more desirable form of lithium doping is Mg$_{1-x}$Li$_x$B$_2$ with the lithium substituting for the
magnesium and no disruption of the boron sublattice.

The samples annealed at 500$^o$C for 1000 hours suggest that at low levels, Li doping has little or no effect on
$H_{c2}$. The $9.50 \times 10^{18}$ cm$^{-2}$ fluence sample has twice the amount of Li of the $4.75 \times
10^{18}$ cm$^{-2}$ fluence sample, yet their $H_{c2}$(T=0) values are approximately equal and only slightly differ
from the un-damaged case. If the 18 T H$_{c2}$ value seen for the $4.75 \times 10^{18}$ cm$^{-2}$ fluence level is
truly a 2 T enhancement relative to the un-damaged sample and is the result of Li doping, then we would expect the
$9.50 \times 10^{18}$ cm$^{-2}$ fluence level to exhibit a $H_{c2}(T=0)$ near 20 T, which we did not observe.
Since we can not insure that we have fully annealed out all of defects, it is possible that this slight increase
in $H_{c2}(T=0)$ is a result of scattering associated with structural defects. In this case $H_{c2}(T=0)$ should
be determined more by $T_c$ than by the particular fluence level or post exposure annealing profile. The $T_c$
values for the $4.75 \times 10^{18}$ cm$^{-2}$ fluence level annealed at 500$^o$C for 24 and 1000 hours and that
of the $9.50 \times 10^{18}$ cm$^{-2}$ fluence level annealed at 500$^o$C for 1000 hours are all within half a
degree of one another. This suggests that $H_{c2}(T=0)$ is controlled more by scattering associated with residual
defects rather than any inadvertent Li doping. Since these low levels of Li do not appear to have any major impact
on the superconducting properties, we can limit the discussion to possible influences of disorder, scattering, and
possible changes in the Fermi surface.

The evolution of the superconducting transition temperature as a function of the unit cell dimensions ($\Delta$a,
$\Delta$c, and V/V$_0$) shows definite trends and is in good agreement with the results of other neutron
irradiation studies \cite{14,20} (Figure 18). Although there is considerable spread in the data, all three reports
show that T$_c$ tends to decrease with both $\Delta$a and $\Delta$c and hence V/V$_0$. These data show that, for
neutron damaged samples, there is some correlation between the unit cell dimensions and the superconducting
properties. They do not, however, uniquely determine if the changes are a result of changes in the Fermi surface,
perhaps due to a repositioning of the atoms, or if they are due to an introduction of additional scattering
centers. In the case of substantial neutron irradiation induced damage, where T$_c$ is below 10 K, NMR
measurements indicate the suppression of superconductivity is the result of a decrease in the density of states of
the boron p$_{x,y}$ orbitals \cite{29}. This technique has been successful in experimentally determining a
decrease in the density of states in Mg$_{1-x}$Al$_x$B$_2$ and AlB$_2$ \cite{30,31}. It should be noted however,
that measurements of the nuclear spin relaxation rate, T$^{-1}_1$, on $^{11}$B yielded comparable T$_1$T values,
which are directly related to the density of states at the Fermi surface, for both a neutron irradiated sample
which had a T$_c$ near 7 K \cite{29} and a sample with 30\% aluminum substituted for magnesium which had a T$_c$
near 25 K \cite{30}. That the neutron damaged sample exhibits a dramatically lower T$_c$ despite having virtually
the same density of states at the Fermi surface suggests additional mechanisms act to suppress superconductivity
in neutron irradiated MgB$_2$. This notion is supported by the fact that for the 24 hour anneals at 300$^o$C,
400$^o$C, and 500$^o$C, the relative changes in the lattice parameters are fairly small. This should only
contribute to minor changes in the Fermi surface, yet T$_c$ is suppressed to near 25 K in the case of the 300$^o$C
anneal.

Although there exist correlations between our data and literature reports on the evolution of T$_c$ and the
lattice parameters, the evolution of $H_{c2}(T=0)$ varies depending upon the irradiation conditions. This present
work on heavily irradiated MgB$_2$ containing natural boron followed by post exposure annealing led to little or
no enhancement of the upper critical field. In contrast others have reported enhancements of $H_{c2}$ for low
fluence levels on MgB$_2$ containing either natural boron or isotopically enriched $^{11}$B \cite{15,18} as well
as fast neutron irradition of natural boron containing MgB$_2$ \cite{1217}.

In order to better understand this conspicuous difference it is useful to review what happened during damage and
subsequent annealing. When a neutron capture and subsequent alpha decay event occurs, the resultant damage is a
large cluster of dislocations. Primary knock-on events cause a cascade of displacements which can be spread over a
distance as large as a hundred atomic distances \cite{24}.

As mentioned previously, $T_c$ suppression in as damaged samples containing natural boron is much more rapid than
those with either isotopically enriched $^{11}$B or those which have been shielded from low energy neutrons.
Therefore it is presumably the clusters associated with the neutron capture event which are primarily responsible
for the suppression of $T_c$. These clusters presumably have low activation energy and can be largely repaired by
the annealing process, which explains why we saw such a rapid increase in $T_c$ for short time and low temperature
anneals. In order to see an enhancement in $H_{c2}$ we need to anneal for a sufficient time and temperature so as
to increase $T_c$ to a level where we aren't being limited by a low transition temperature. Since the defect
clusters from the alpha decay of $^{11}$B to $^7$Li are presumably large, and we are relying on random diffusion
to recombine vacancies and interstitials, higher reaction temperatures are necessary to restore $T_c$.

By going to higher annealing temperatures we begin to repair defects with higher activation energies. Therefore,
regardless of the actual energy associated with repairing the defects responsible for the enhancement of $H_{c2}$,
we are most likely annealing many of them away while restoring $T_c$.

If it is these defects which are responsible for $H_{c2}$ enhancement then it is the annealing process which
causes the different development of $H_{c2}$ we observe.

Based on this analysis it seems likely that a study of wire samples that have been exposed to much smaller
fluences of neutrons (e.g. an exposure level that yields as damaged samples with a transition temperature in the
range 30 $< T_c <$ 37) should yield different, and higher, $H_{c2}(T)$ data.

The transmutation of $^{10}$B following the absorption of a neutron results in the emission of an alpha particle
of energy 1.7 MeV, which is quite comparable in energy to the 2 MeV He ion irradiation of MgB$_2$ films done by
Gandikota et al. \cite{476,477}. One might therefore expect rather similar results between the two studies.
Comparison of the results indicates these two different types of irradiation give rise to different
superconducting and normal state properties. Gandikota and coworkers irradiated with heavy ion fluences up to $1.2
\times 10^{17}$ cm$^{-2}$, which was sufficient to suppress $T_c$ to below 5 K \cite{476}. Unlike the case of
heavy neutron irradiation presented here, He$^{++}$ irradiation led to a slight enhancement in $H_{c2}(T=0)$ for
samples exposed to lower fluence levels, where $T_c$ was only suppressed by a few degrees \cite{477}, analogous to
the results obtained for lower fluence level neutron irradiation \cite{15,18}. $H_{c2}(T=0)$ did tend to scale
with $T_c$ for samples with $T_c$ below 30 K and, as with neutron irradiated samples, $H_{c2}(T)$ begins to
exhibit more WHH like behavior when $T_c$ in the vicinity of 10 K. Additionally, $T_c$ could readily be restored
by annealing at temperatures as low as 100$^o$C \cite{477}. However, x-ray measurements indicated that the He ion
irradiation did not change the c-lattice parameter, even for the highest fluence level \cite{476}. Thus, He$^{++}$
irradiation presumably creates more point defects rather than large defect clusters which are likely to arise from
neutron irradiation. He ion irradiation therefore presents yet another different avenue to systematically change
the scattering levels in MgB$_2$.

Damage induced by neutron irradiation is fundamentally different from doping with aluminum or carbon. Both carbon
and aluminum enter the structure and are believed to act as point defects in addition to electron doping the
system. Arguments based on the temperature dependence of the anisotropy of the upper critical field suggest both
carbon and aluminum doping increase scattering within the $\pi$ band relative to the $\sigma$ band, with the
effect is much more pronounced in the case of carbon doping \cite{7}. It should be noted that other researchers
have concluded that changes in the temperature and magnetic field dependence of the thermal conductivity,
$\kappa$(T,H), for carbon substituted single crystals are consistent with carbon doping resulting in an
enhancement in intra-$\sigma$-band scattering \cite{1226}. Although there exists some debate as to the exact
nature of the enhancement in scattering in carbon doped MgB$_2$, it is clear that the development of H$_{c2}$ is
dominated by scattering effects for carbon substitutions and and Fermi surface changes effects for aluminum
substitution. For neutron irradiated samples, as discussed above, the effects of the introduction of Li through
nuclear processes are negligible and the changes in the superconducting properties are due to an increase in
interband scattering and a decrease in the density of states. Insights can be made into the effects of neutron
irradiated samples by comparison to carbon and aluminum doping.

The different development of interdependencies of H$^{\parallel ab} _{c2}$, T$_c$, and $\rho_0$ in aluminum doped
\cite{9}, carbon doped \cite{187,188}, and neutron irradiated samples, those exposed to a fluence of $4.75 \times
10^{18}$ cm$^{-2}$ and annealed for 24 hours, are plotted in figure 19. Figure 19a shows the evolution of the
transition temperature as a function of residual resistivity for these three types of perturbations. The
suppression of $T_c$ with $\rho_0$ for low resistivity values is quite similar for the neutron damaged and
aluminum doped samples; $T_c$ drops rapidly with increased scattering. In contrast, the suppression of $T_c$ as a
function of resistivity in carbon doped samples is quite gradual. This is consistent with both aluminum doping and
neutron damaged samples having more interband scattering than carbon doped compounds.

Figure 19b plots $H_{c2}(T=0)$ as a function of $\rho_0$. Here all three perturbations behave uniquely. The
aluminum and neutron damage samples show a decrease in $H_{c2}$, whereas the carbon doped samples show a dramatic
increases. The decrease in $H_{c2}$ is more rapid in the aluminum doped samples than in neutron damaged samples.
This can also be seen by plotting $H_{c2}$ as a function of $T_c$ (Figure 19c).

Direct comparison between the evolution of $T_c$ and $H_{c2}(T=0)$ for carbon doping and neutron irradiation shows
the scattering associated with each type of perturbation is different. Within the model proposed by Gurevich
\cite{11}, enhancements in $H_{c2}$ result from differences in the relative strength of scattering within each
band, whereas the suppression of $T_c$ is a function of scattering between the bands. The neutron irradiated
samples on these plots with a $T_c$ greater than 25 K are samples which were annealed at 300$^o$C, 400$^o$C, and
500$^o$C. In each of these samples, annealing has reduced the a-lattice parameter to nearly the undamaged value
(Figure 2b). With the defect structure lying between nearly undamaged boron planes, one would not expect the
scattering within the $\sigma$ band to be substantially affected. If the scattering was confined primarily to the
3-D $\pi$ band, the resultant differences in intraband diffusivity values should manifest themselves in terms of
enhanced $H_{c2}$. Since no significant enhancement is seen, we believe the scattering is primarily interband
scattering and contributes to the suppression of $T_c$. The range in $T_c$ for $V/V_0 \sim 1$ in post exposure
annealed samples (Figure 18c) supports the notion that neutron irradiation increases interband scattering.

The suppression in $T_c$ in neutron irradiated samples is the result of a combination of a decrease in the density
of states at the Fermi surface and an increase in interband scattering. In the case of aluminum doping, the
suppression of $T_c$ is believed to be primarily a result of a changes in the Fermi surface \cite{7}. Samples of
neutron damaged MgB$_2$ exhibit higher $H_{c2}$ values than aluminum doped MgB$_2$ samples with similar $T_c$
values (Figure 19c). This suggests that, for neutron irradiated samples, changes in the density of states suppress
$H_{c2}$ to a greater degree than interband scattering.

\section{Conclusions}

We systematically studied the effects of neutron fluence level, annealing temperature, and annealing time on the
superconducting and normal state properties of MgB$_2$. As damaged samples showed an anisotropic expansion of the
unit cell and a suppression of $T_c$ to below 5 K. Defects introduced by the irradiation process had relatively
low activation energy and hence much of the damage could be repaired by post-exposure annealing. $H_{c2}$ values
tended to scale with $T_c$ and the evolution of $H_{c2}$ with $\rho _0$ was dramatically different than in the
case of carbon doping. Lithium produced through nuclear processes appeared to have no significant effects on the
superconducting properties. We attribute the changes in the superconducting properties of neutron irradiated
samples primarily to an increase in scattering between the two bands and possible changes in the density of state
at the Fermi surface.

\section{Acknowledgements}

Ames Laboratory is operated for the U.S. Department of Energy by Iowa State University under Contract No.
W-7405-Eng-82. This work was supported by the Director for Energy Research, Office of Basic Energy Sciences. A
portion of this work was performed at the National High Magnetic Field Laboratory, which is supported by NSF
Cooperative Agreement No. DMR-0084173 and by the State of Florida.

\clearpage

\begin{figure}
\begin{center}
\includegraphics[angle=0,width=180mm]{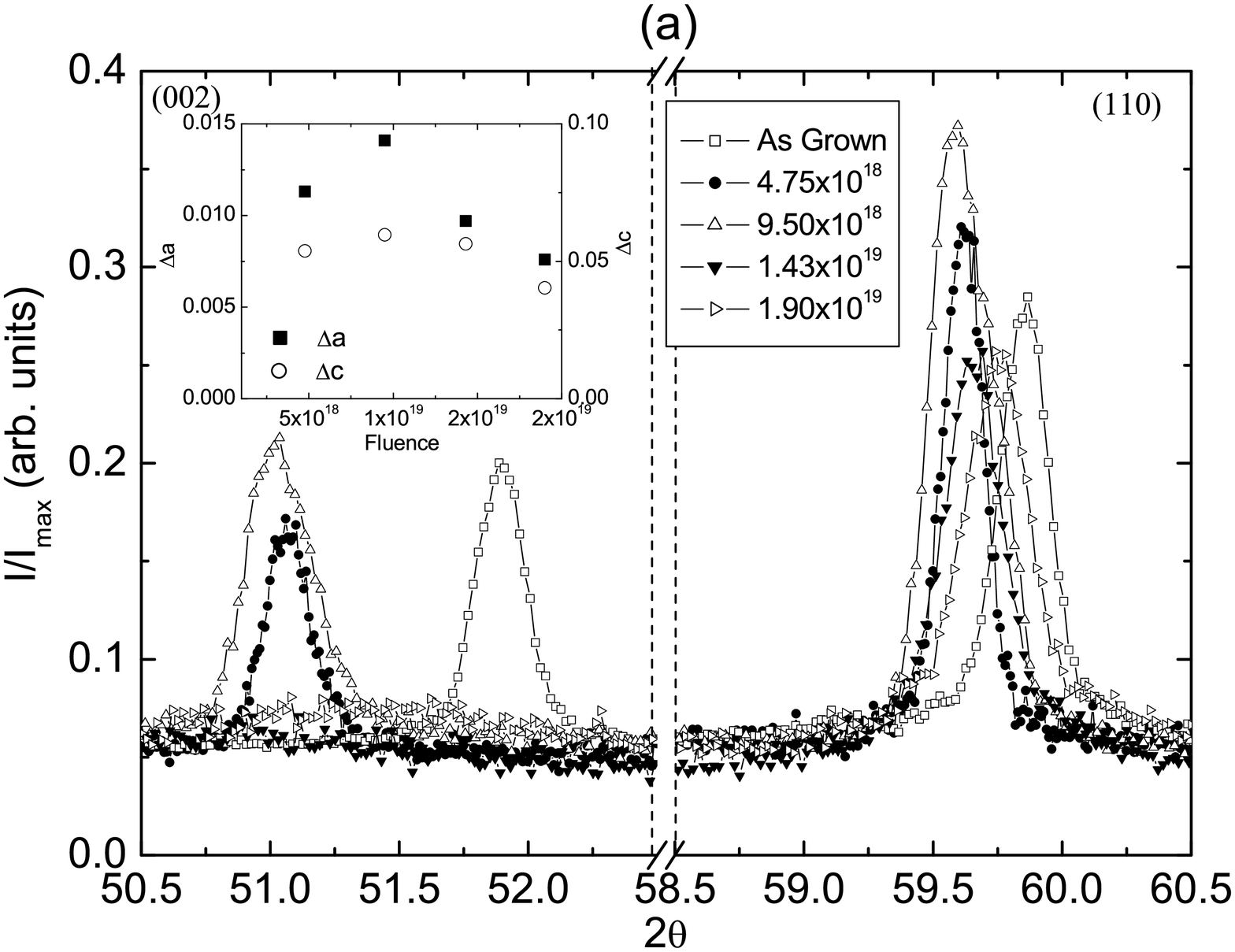}
\end{center}
\end{figure}

\clearpage

\begin{figure}
\begin{center}
\includegraphics[angle=0,width=180mm]{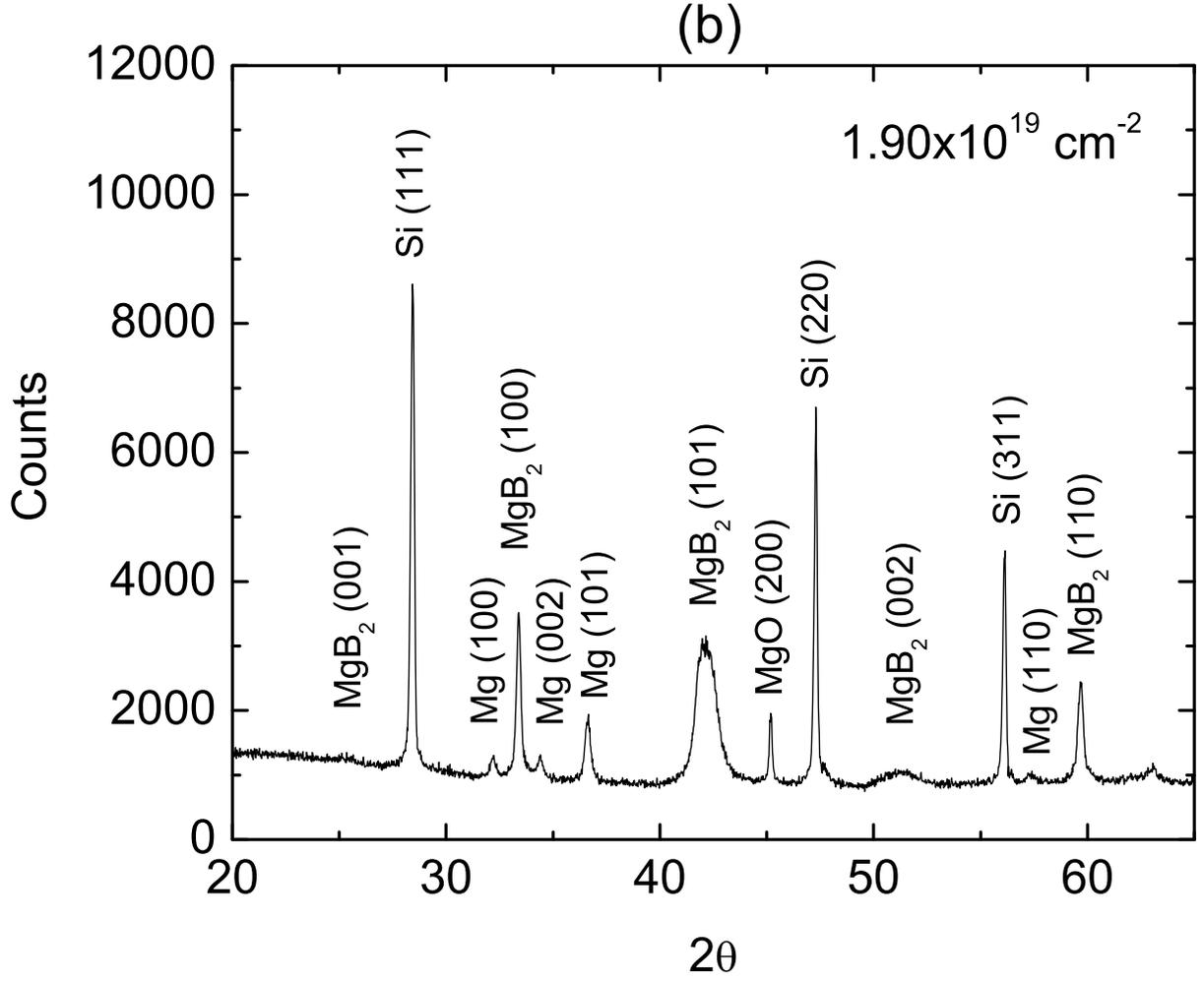}
\end{center}
\caption{(a) (002) and (110) X-ray peaks for un-damaged, $4.75 \times 10^{18}$, $9.50 \times 10^{18}$, $1.43
\times 10^{19}$, and $1.90 \times 10^{19}$ cm$^{-2}$ fluence as-damaged samples. The highest two fluence levels
show a broadened (002) peak indicating a decrease in order along the c-axis. The inset shows the relative shift of
the a- and c- lattice parameters as a function of fluence level. (b) The fuller diffraction pattern of the
as-damaged $1.90 \times 10^{19}$ cm$^{-2}$ fluence level sample. On this fuller range the broadened (002) peak is
clearly visible.}\label{f1}
\end{figure}

\clearpage

\begin{figure}
\begin{center}
\includegraphics[angle=0,width=180mm]{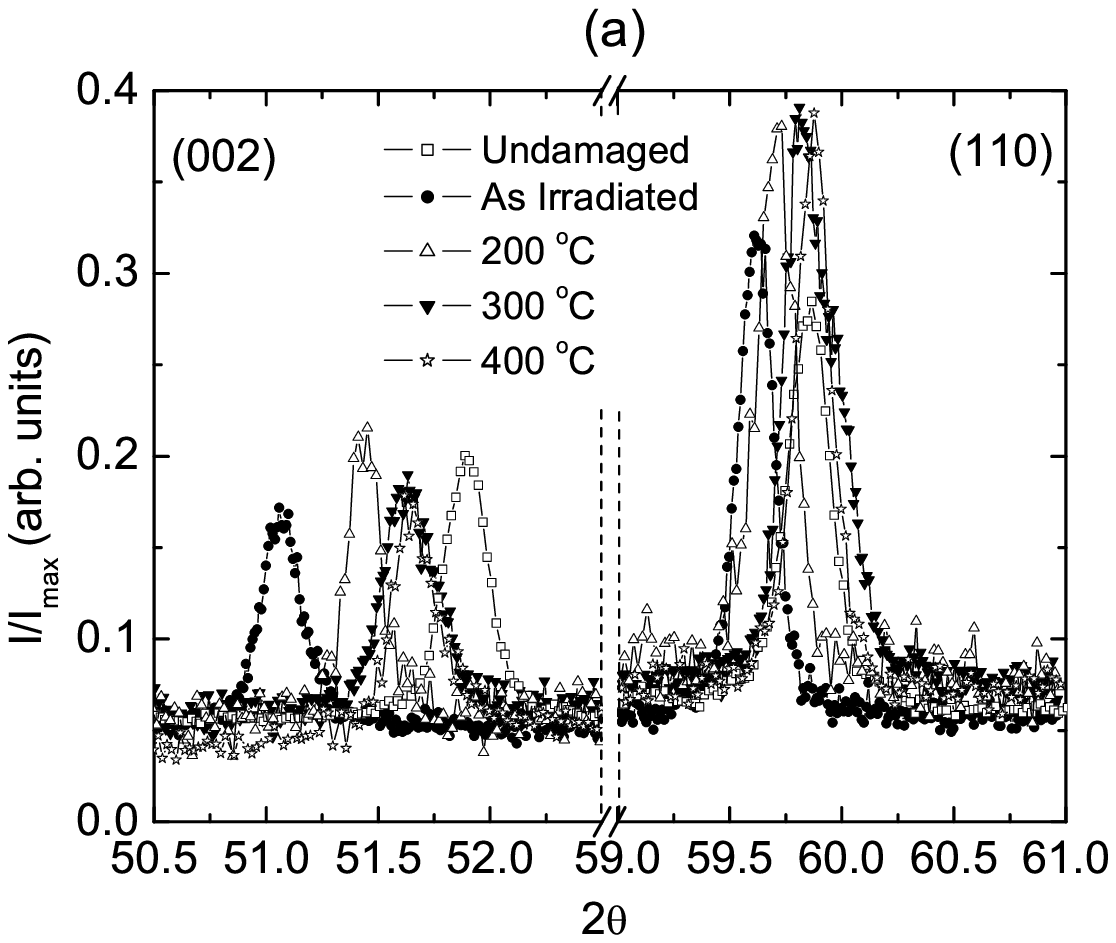}
\end{center}
\end{figure}

\clearpage

\begin{figure}
\begin{center}
\includegraphics[angle=0,width=180mm]{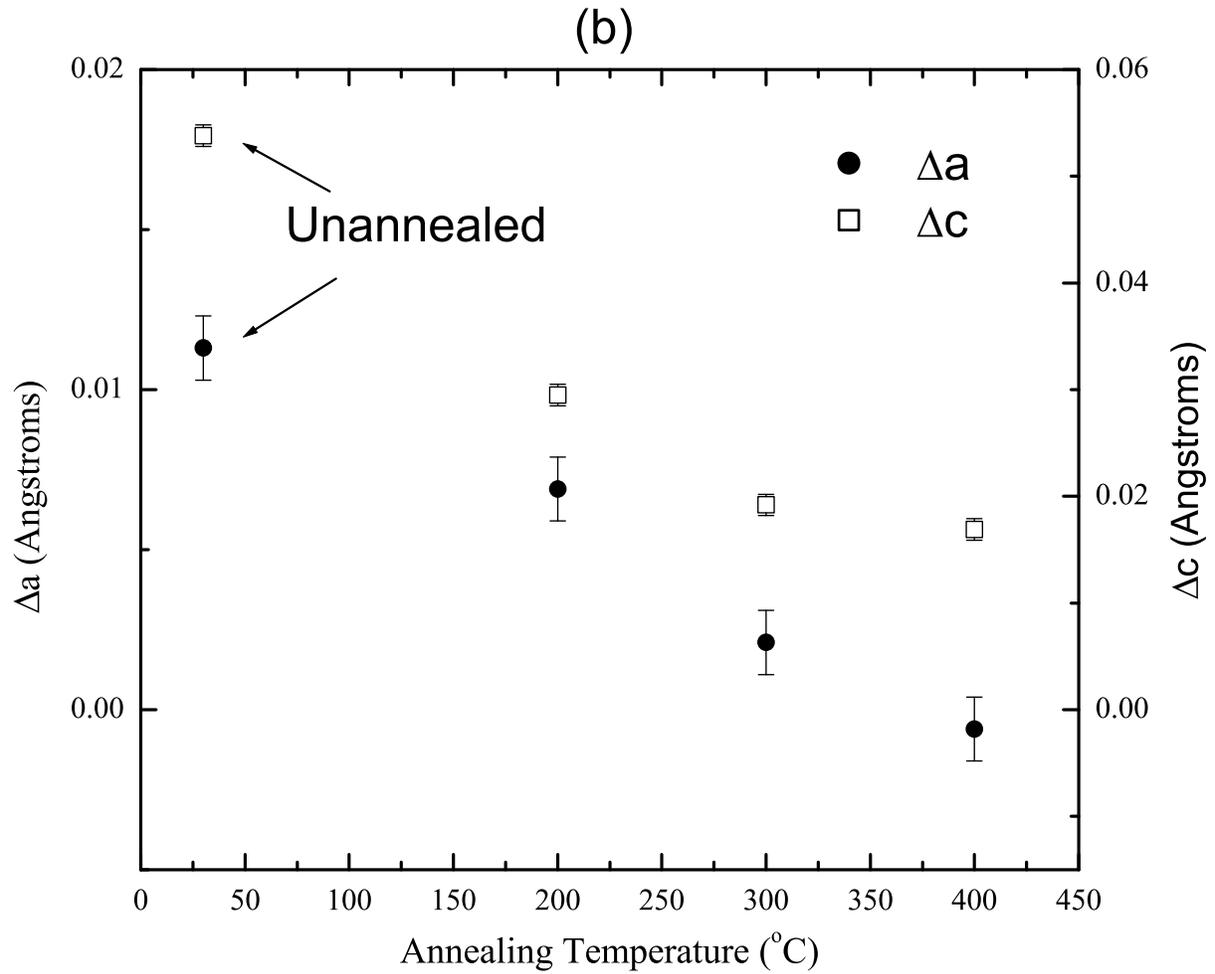}
\end{center}
\caption{(a) (002) and (110) x-ray peaks used to determine the a- and c-lattice parameters for the set of 24 hour
anneals on samples exposed to a fluence of $4.75 \times 10^{18}$ cm$^{-2}$. (b) The evolution of the lattice
parameters as a function of annealing temperature. Closed symbols represent $\Delta$a, and open symbols are
$\Delta$c.}\label{f2}
\end{figure}

\clearpage

\begin{figure}
\begin{center}
\includegraphics[angle=0,width=180mm]{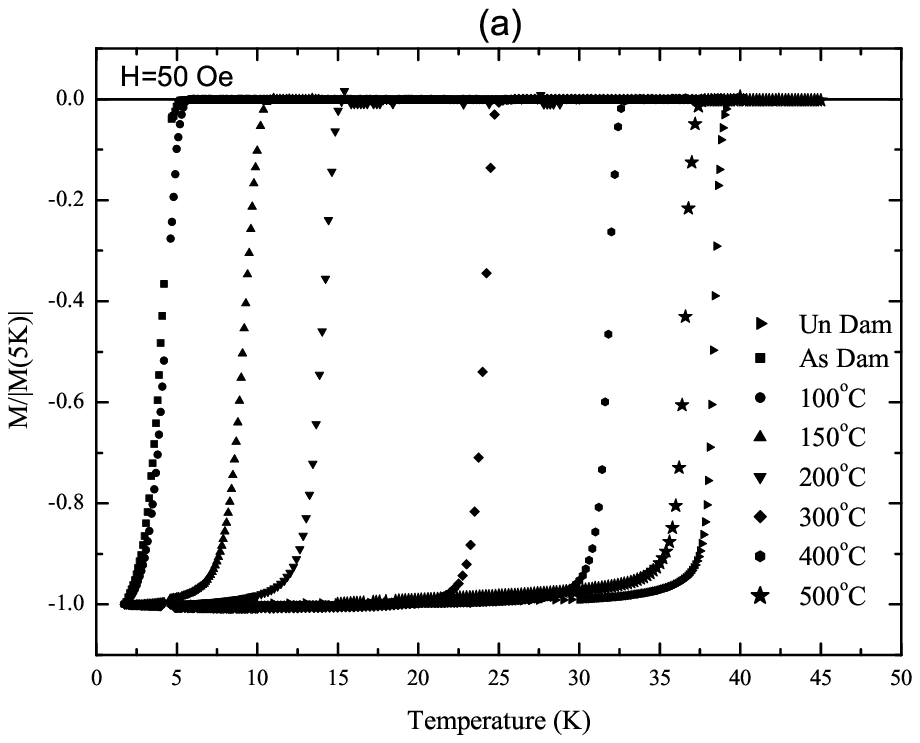}
\end{center}
\end{figure}

\clearpage

\begin{figure}
\begin{center}
\includegraphics[angle=0,width=180mm]{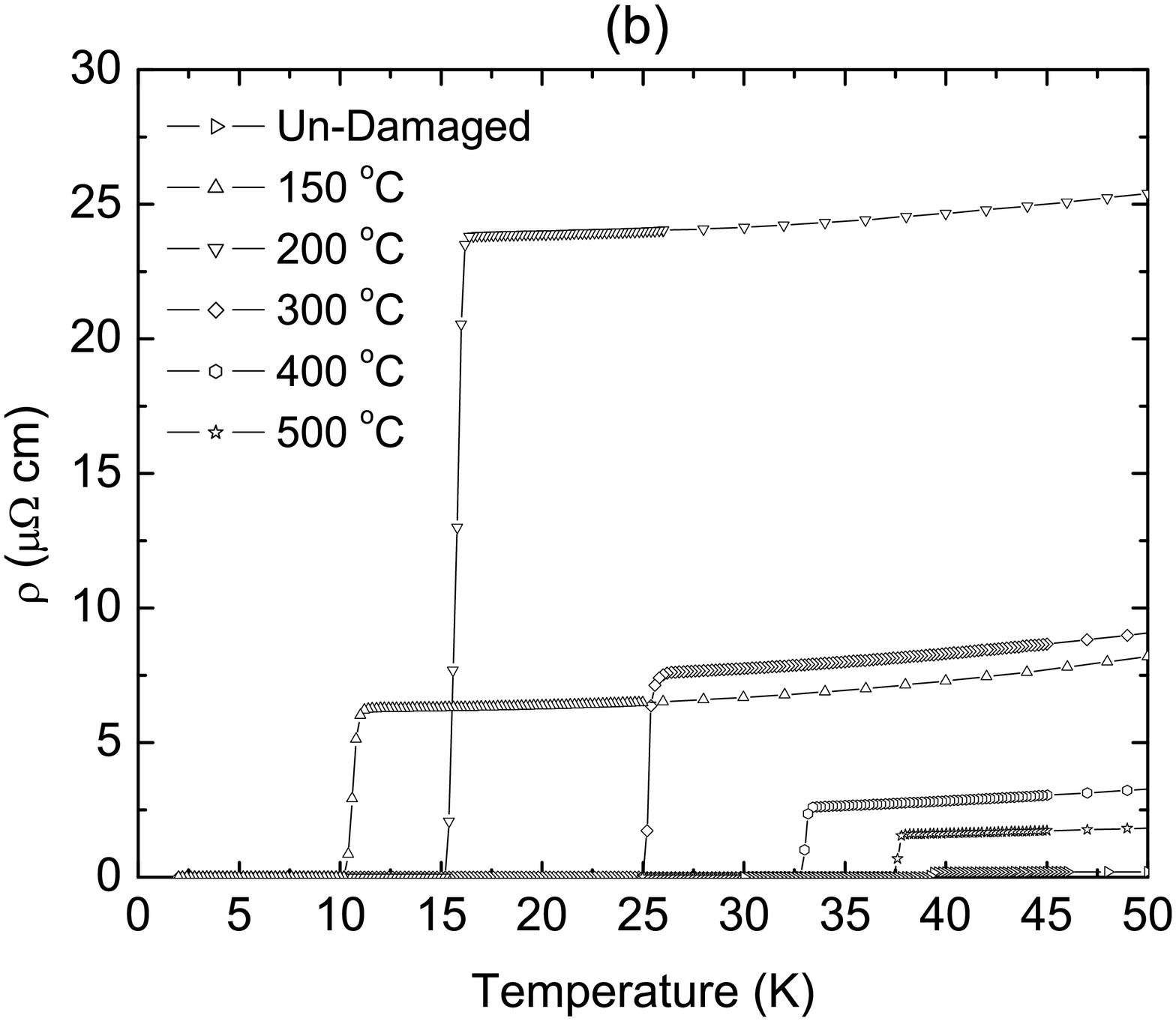}
\end{center}
\end{figure}

\clearpage

\begin{figure}
\begin{center}
\includegraphics[angle=0,width=180mm]{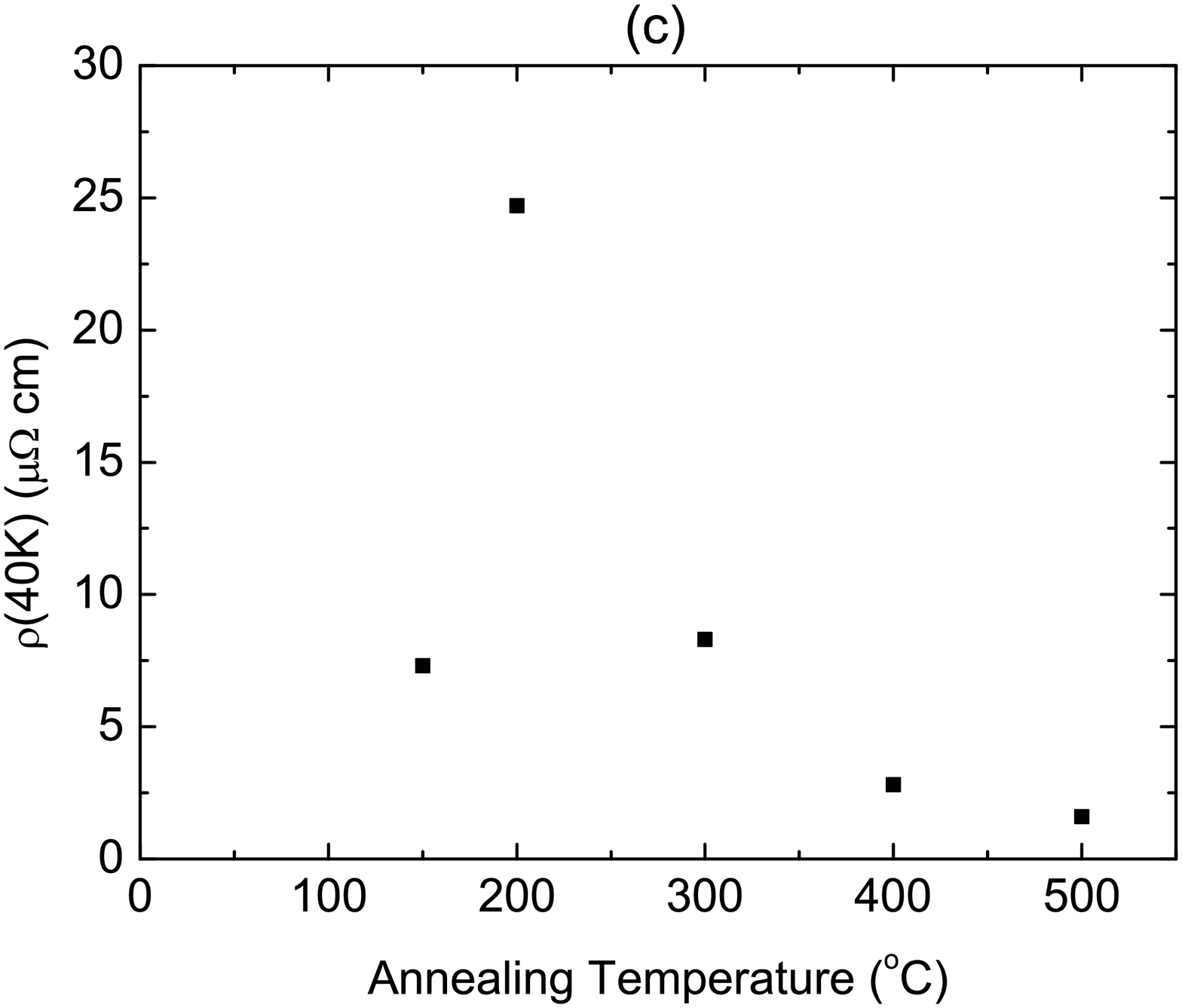}
\end{center}
\end{figure}

\clearpage

\begin{figure}
\begin{center}
\includegraphics[angle=0,width=180mm]{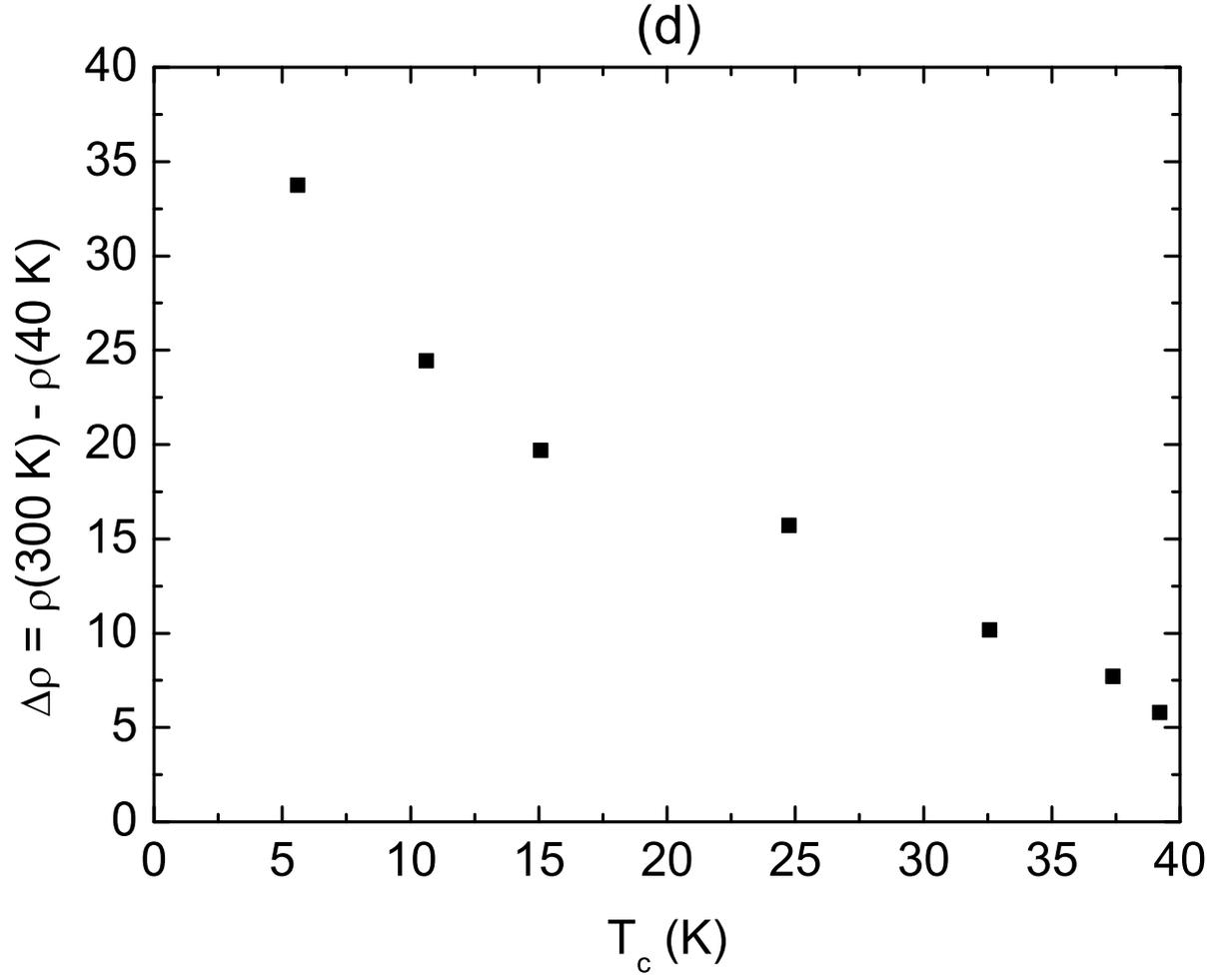}
\end{center}
\caption{(a) Normalized magnetization and (b) resistivity curves for the set of 24 hour anneals on samples exposed
to a fluence of $4.75 \times 10^{18}$ cm$^{-2}$. (c) The normal state resistivity at 40 K. The resistivity shows a
sharp increase at an annealing temperature of 200 $^o$C, then decreases approximately exponentially as a function
of annealing temperature. (d) $\Delta$$\rho$ versus transition temperature.  }\label{f3}
\end{figure}

\clearpage

\begin{figure}
\begin{center}
\includegraphics[angle=0,width=180mm]{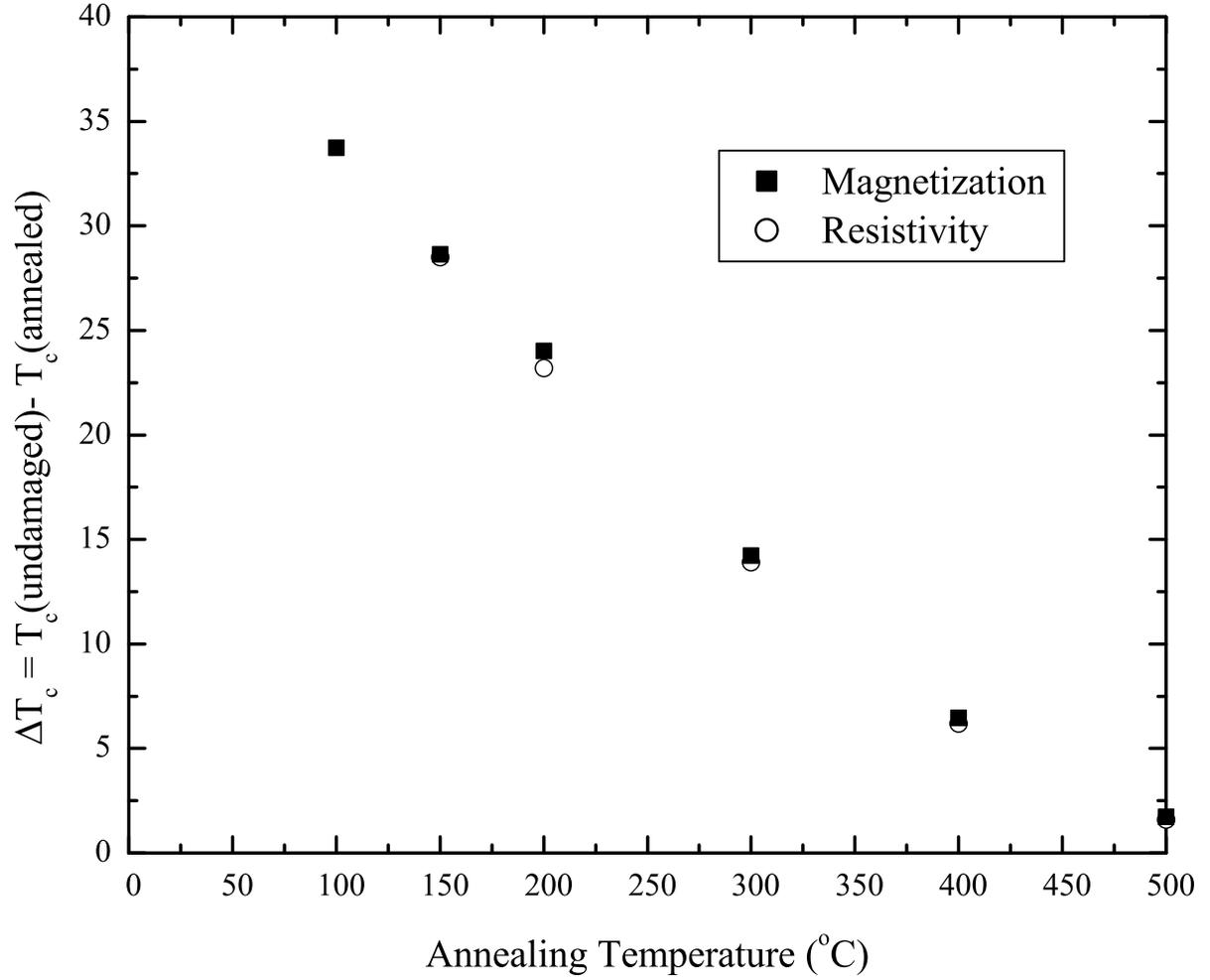}
\end{center}
\caption{Restoration of superconductivity by post exposure annealing for samples exposed to a fluence of $4.75
\times 10^{18}$ cm$^{-2}$ and subsequently annealed for 24 hours at various temperatures. Transition temperatures
were determined using a 1\% screening criteria in magnetization and an onset criteria in resistivity. Annealing at
500$^o$C yields a T$_c$ near 37.5 K, less than 2 K below that of the undamaged sample.}\label{f4}
\end{figure}

\clearpage

\begin{figure}
\begin{center}
\includegraphics[angle=0,width=180mm]{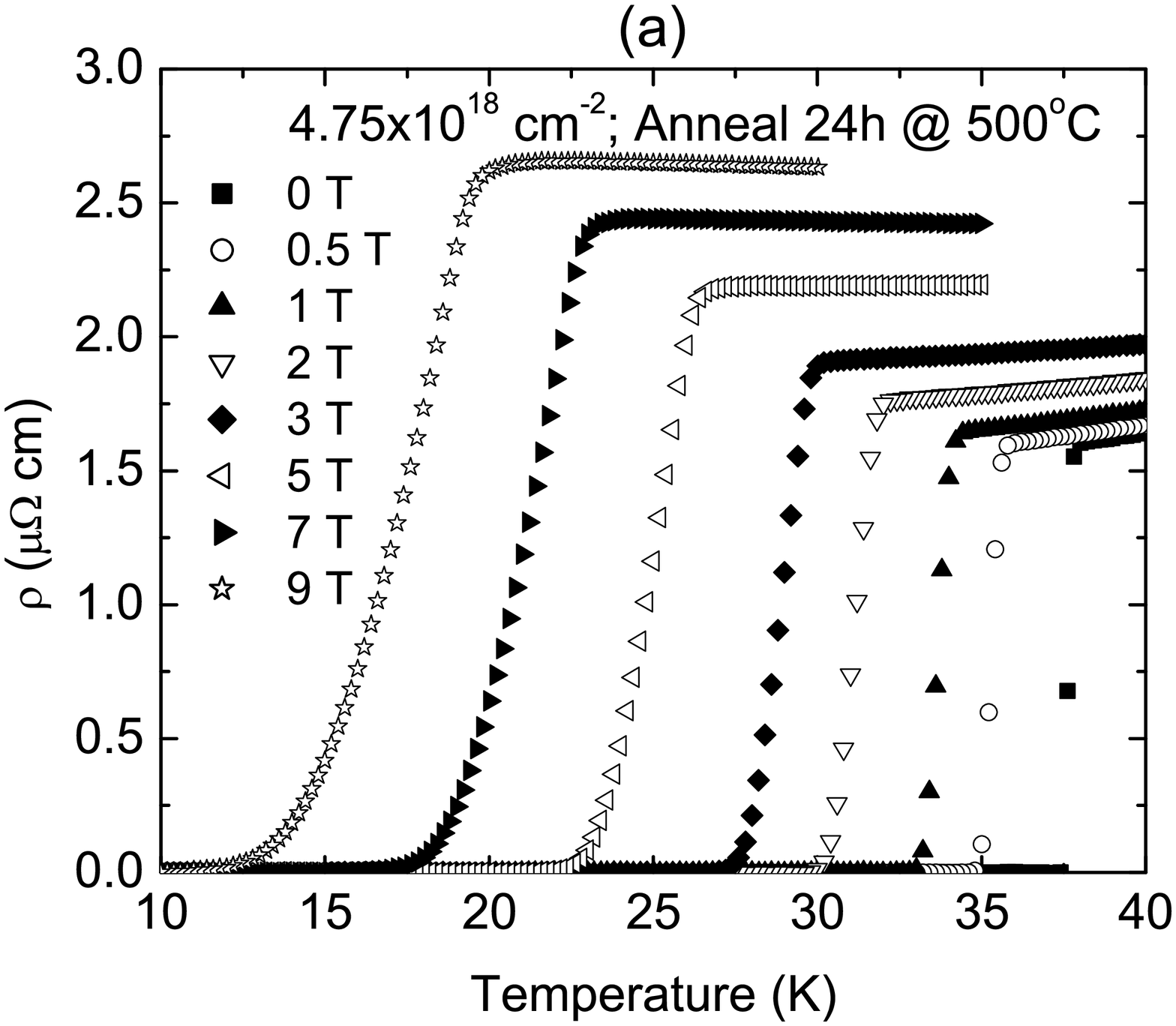}
\end{center}
\end{figure}

\clearpage

\begin{figure}
\begin{center}
\includegraphics[angle=0,width=180mm]{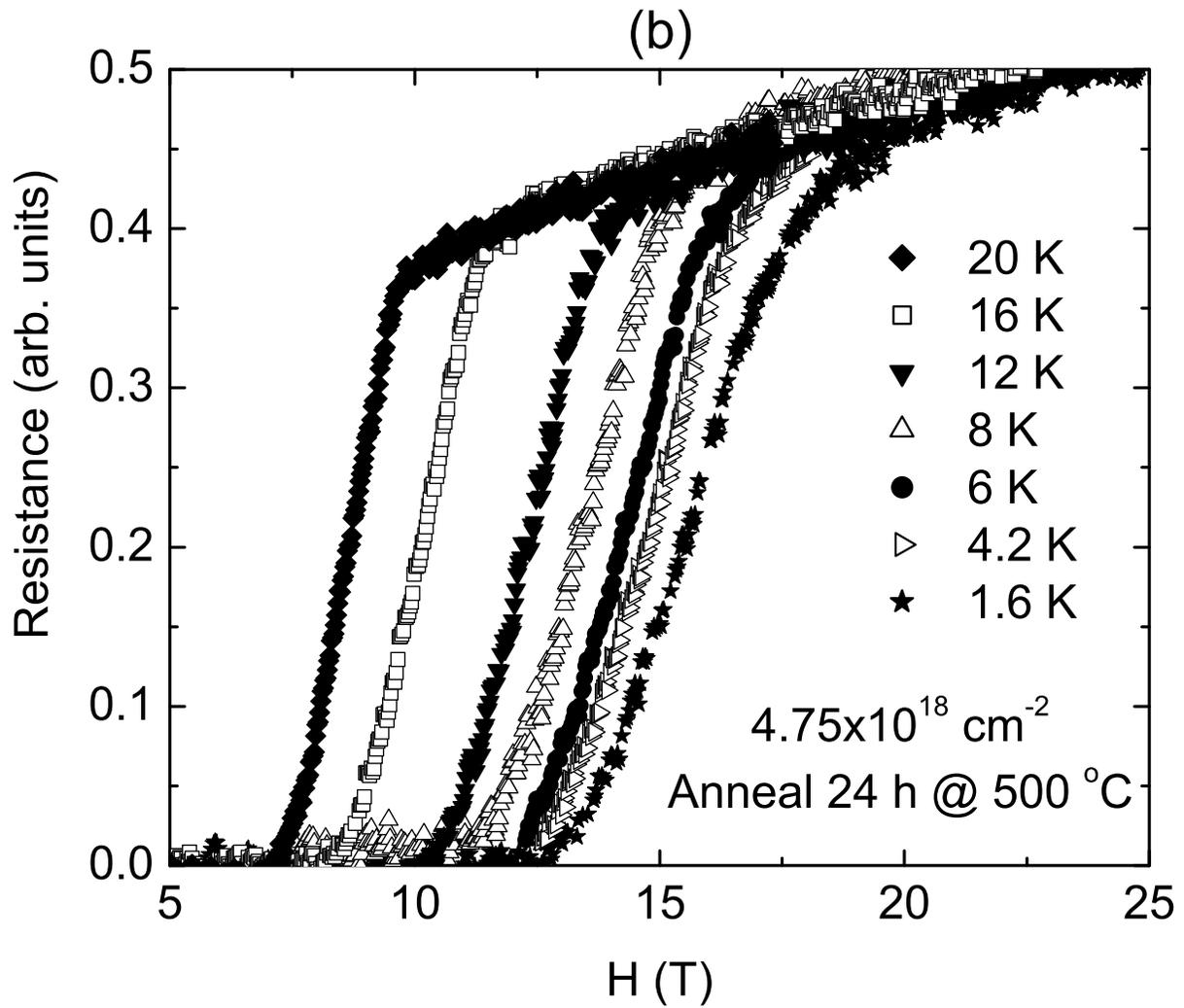}
\end{center}
\caption{Transport measurements to determine upper critical field. (a) Resistivity versus temperature and (b)
resistance versus field. $H_{c2}$ values were determined using the onset criteria in both measurements.
}\label{f5}
\end{figure}

\clearpage

\begin{figure}
\begin{center}
\includegraphics[angle=0,width=180mm]{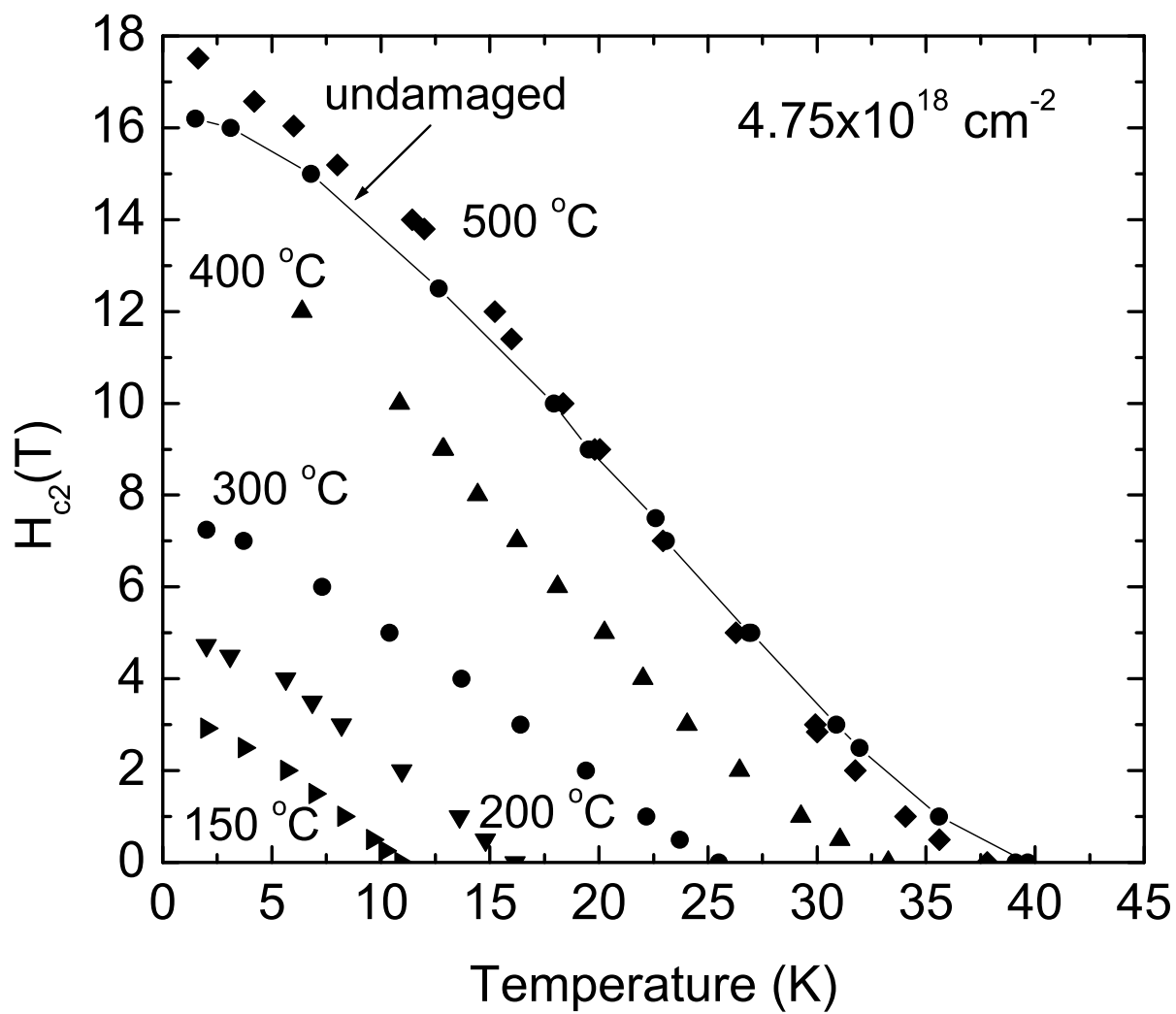}
\end{center}
\caption{Upper critical field curves for undamaged as well as for samples exposed to a fluence of $4.75 \times
10^{18}$ cm$^{-2}$ and annealed at 150$^o$C, 200$^o$C, 300$^o$C, 400$^o$C, and 500$^o$C for 24 hours. }\label{f6}
\end{figure}

\clearpage

\begin{figure}
\begin{center}
\includegraphics[angle=0,width=180mm]{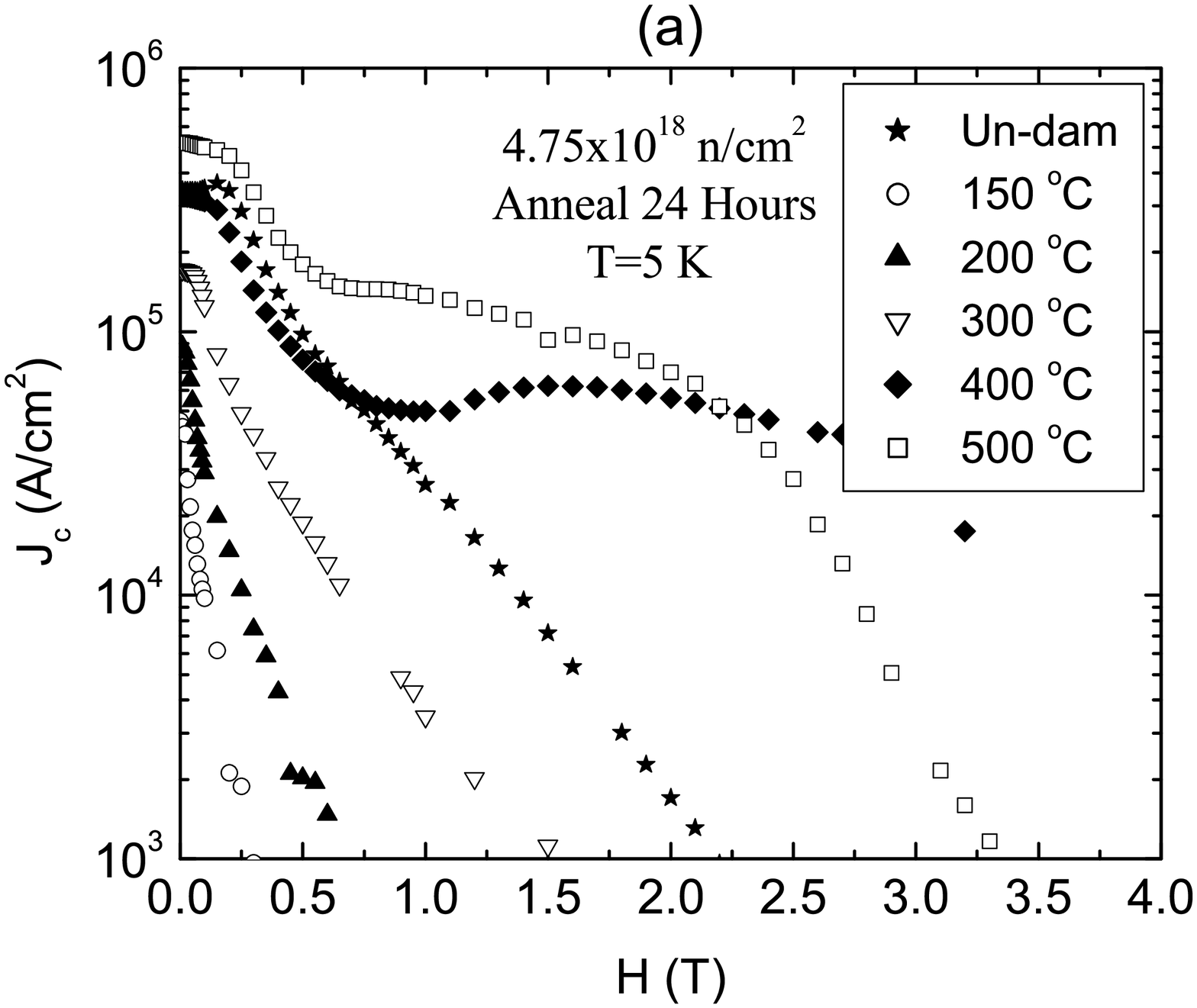}
\end{center}
\end{figure}

\clearpage

\begin{figure}
\begin{center}
\includegraphics[angle=0,width=180mm]{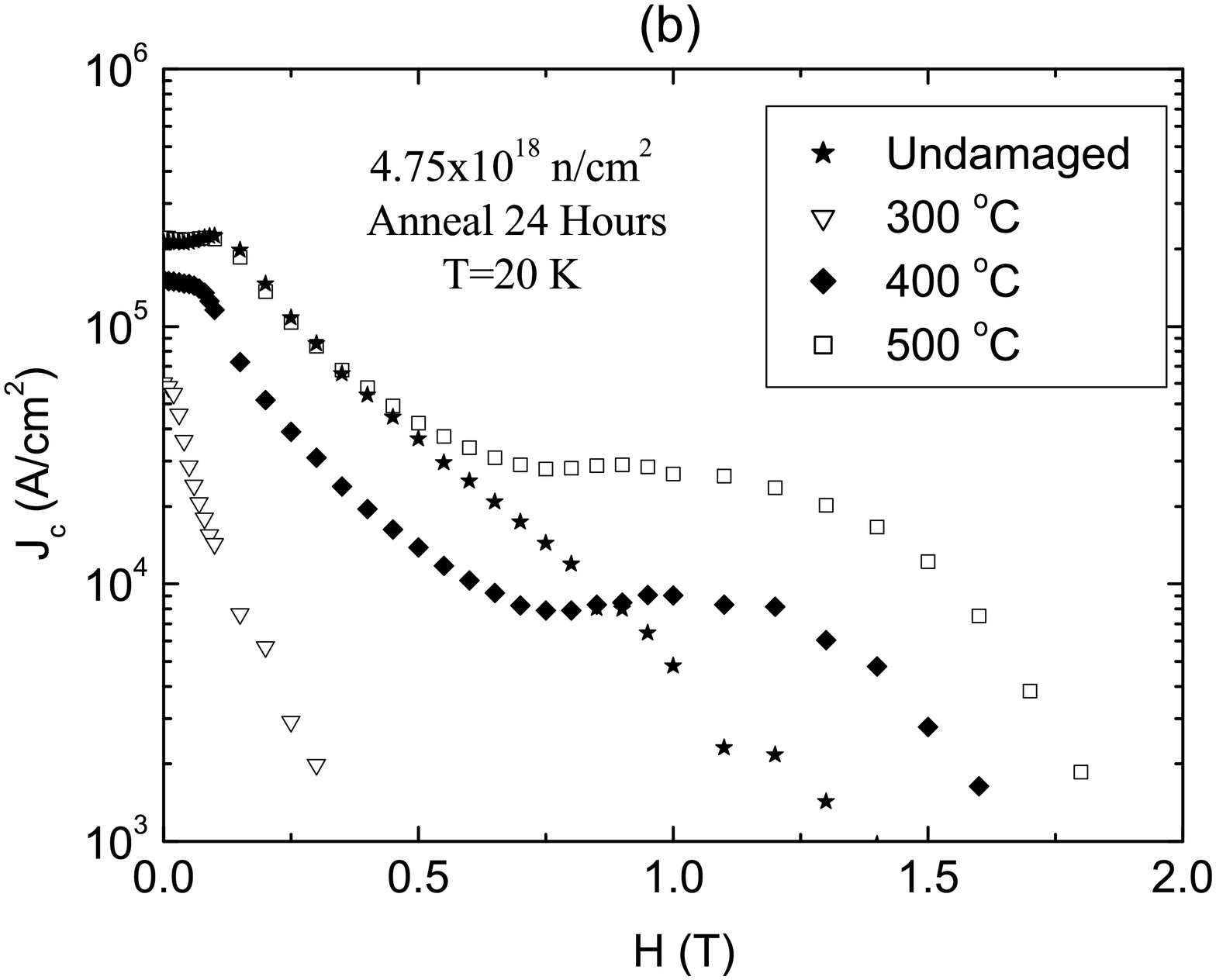}
\end{center}
\end{figure}

\clearpage

\begin{figure}
\begin{center}
\includegraphics[angle=0,width=180mm]{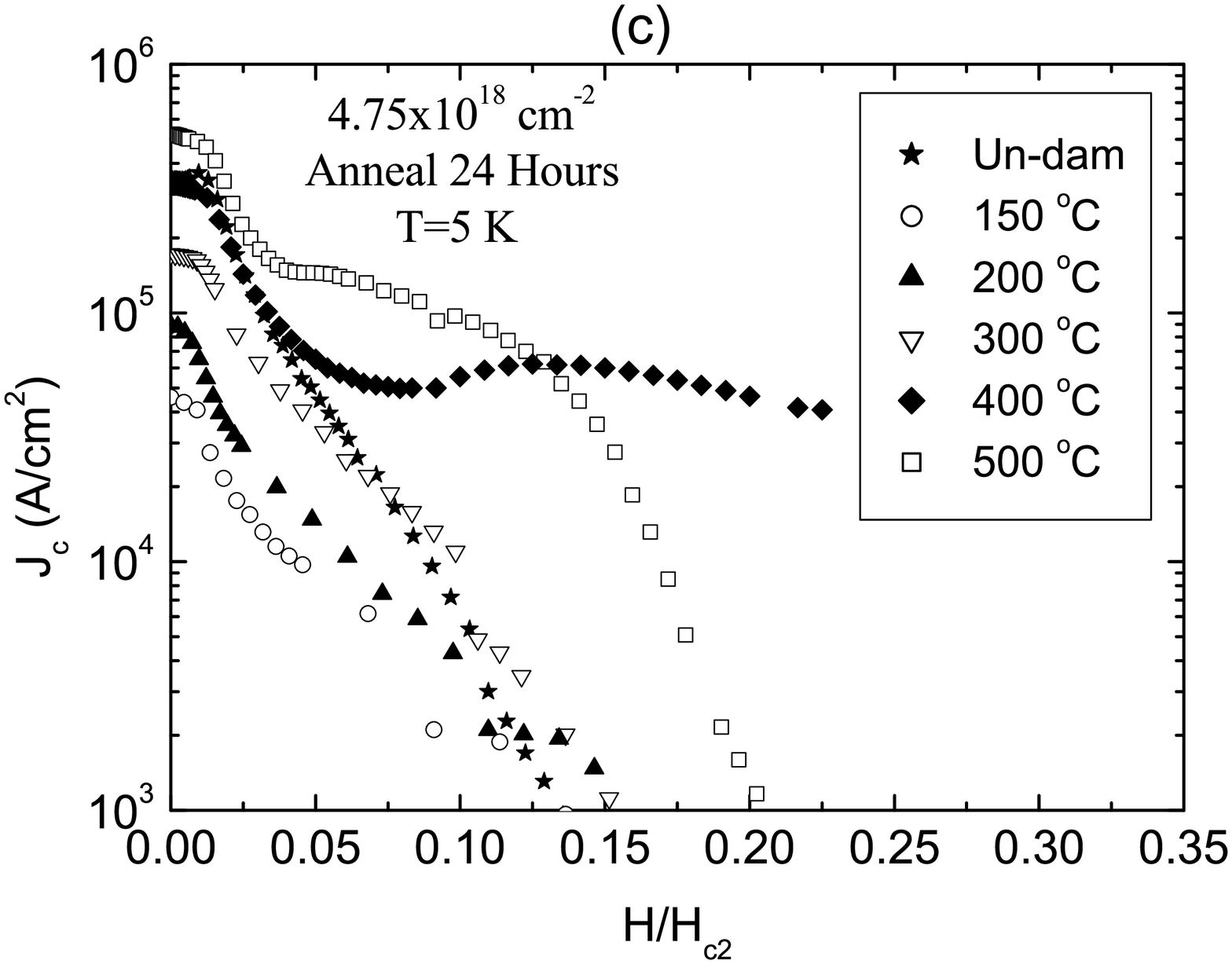}
\end{center}
\end{figure}

\clearpage

\begin{figure}
\begin{center}
\includegraphics[angle=0,width=180mm]{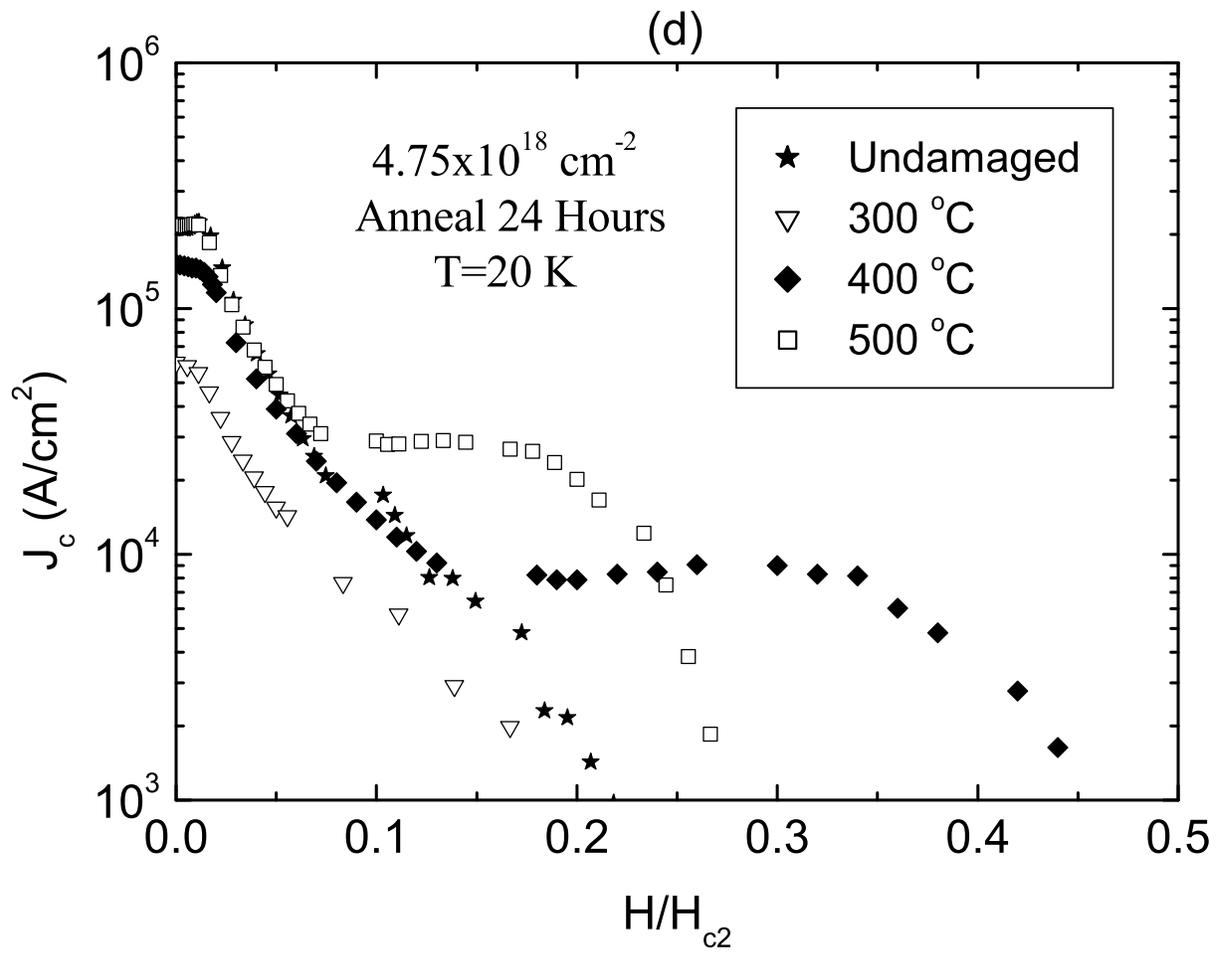}
\end{center}
\caption{Critical current densities as a function of field at (a) 5 K and (b) 20 K for samples irradiated with a
fluence of $4.75 \times 10^{18}$ cm$^{-2}$ and subsequently annealed for 24 hours at various temperatures. Figures
7c and 7d plot $J_c$ as a function of the reduced field $H/H_{c2}$. }\label{f7}
\end{figure}

\clearpage

\begin{figure}
\begin{center}
\includegraphics[angle=0,width=180mm]{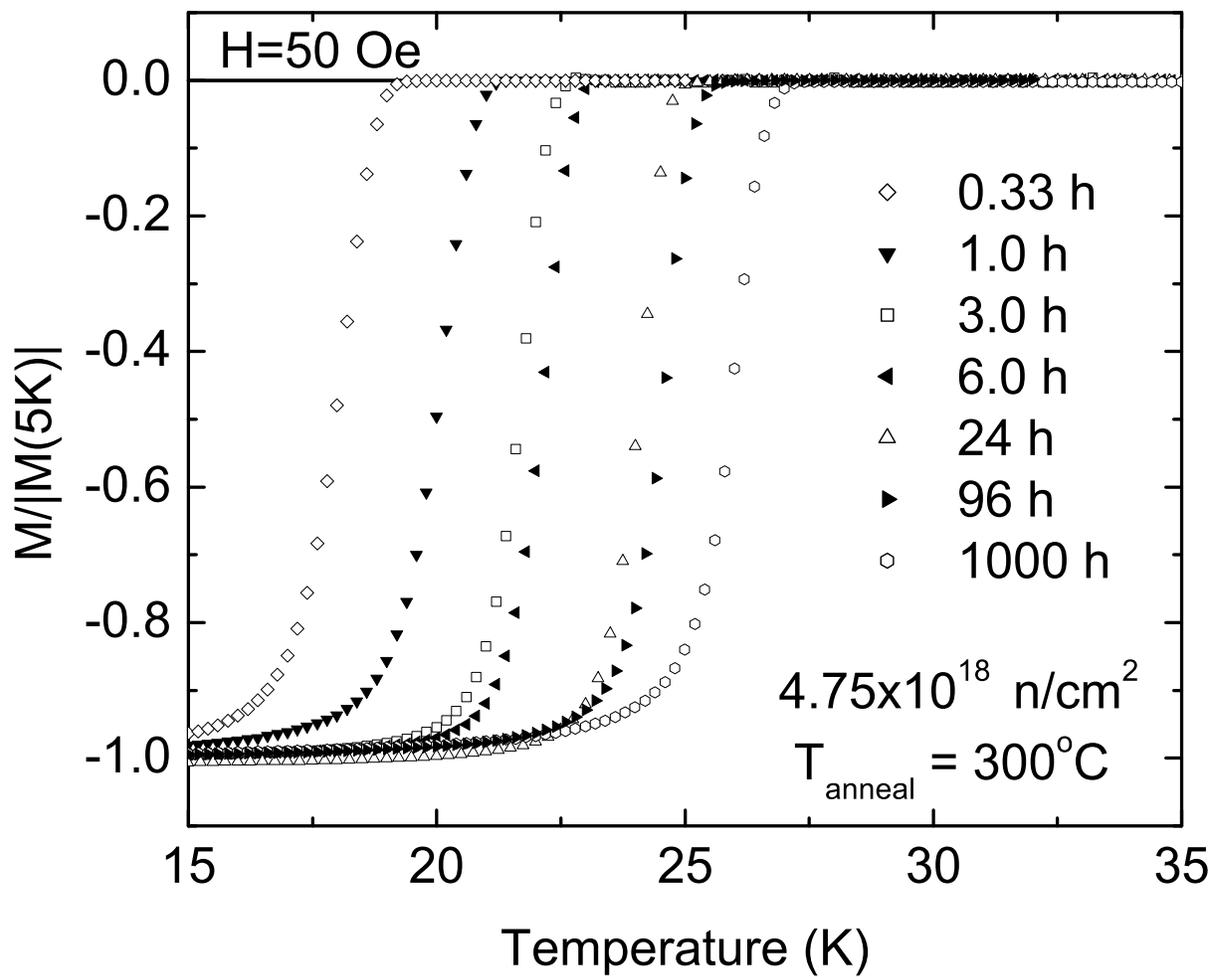}
\end{center}
\caption{Normalized magnetization curves for fluence level of $4.75 \times 10^{18}$ cm$^{-2}$ annealed at 300$^o$C
for 1/3, 1, 3, 6, 24, 96, and 1000 hours.}\label{f8}
\end{figure}

\clearpage

\begin{figure}
\begin{center}
\includegraphics[angle=0,width=180mm]{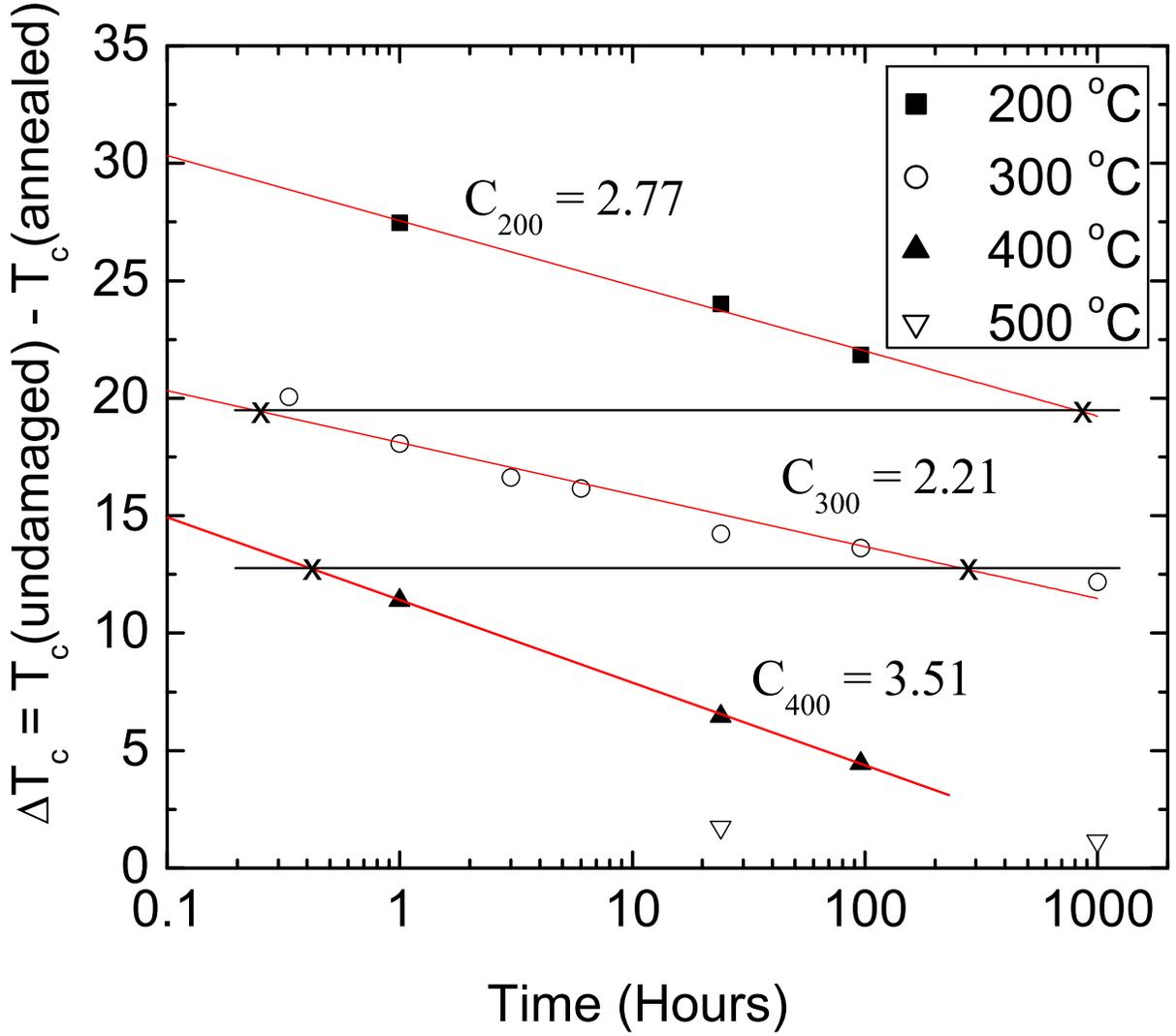}
\end{center}
\caption{(Color online) Semi-log plot of the change of the superconducting transition temperature as a function of
annealing time at various annealing temperatures. All samples shown were exposed to a fluence level of $4.75
\times 10^{18}$ cm$^{-2}$. Using a linear fit over up to three decades yields non-systematic values in the rate
constants for the 200$^o$C, 300$^o$C, and 400$^o$C annealing temperatures. The activation energy is estimated
using the cross cut procedure on extrapolations of these fits to the data (comparison points are given by the x
symbols). By comparing points with identical $\Delta T_c$ values, through equation 3 we obtain estimates of
$E_a$=1.90 eV and 2.15 eV (see equation 4) .}\label{f9}
\end{figure}

\clearpage

\begin{figure}
\begin{center}
\includegraphics[angle=0,width=180mm]{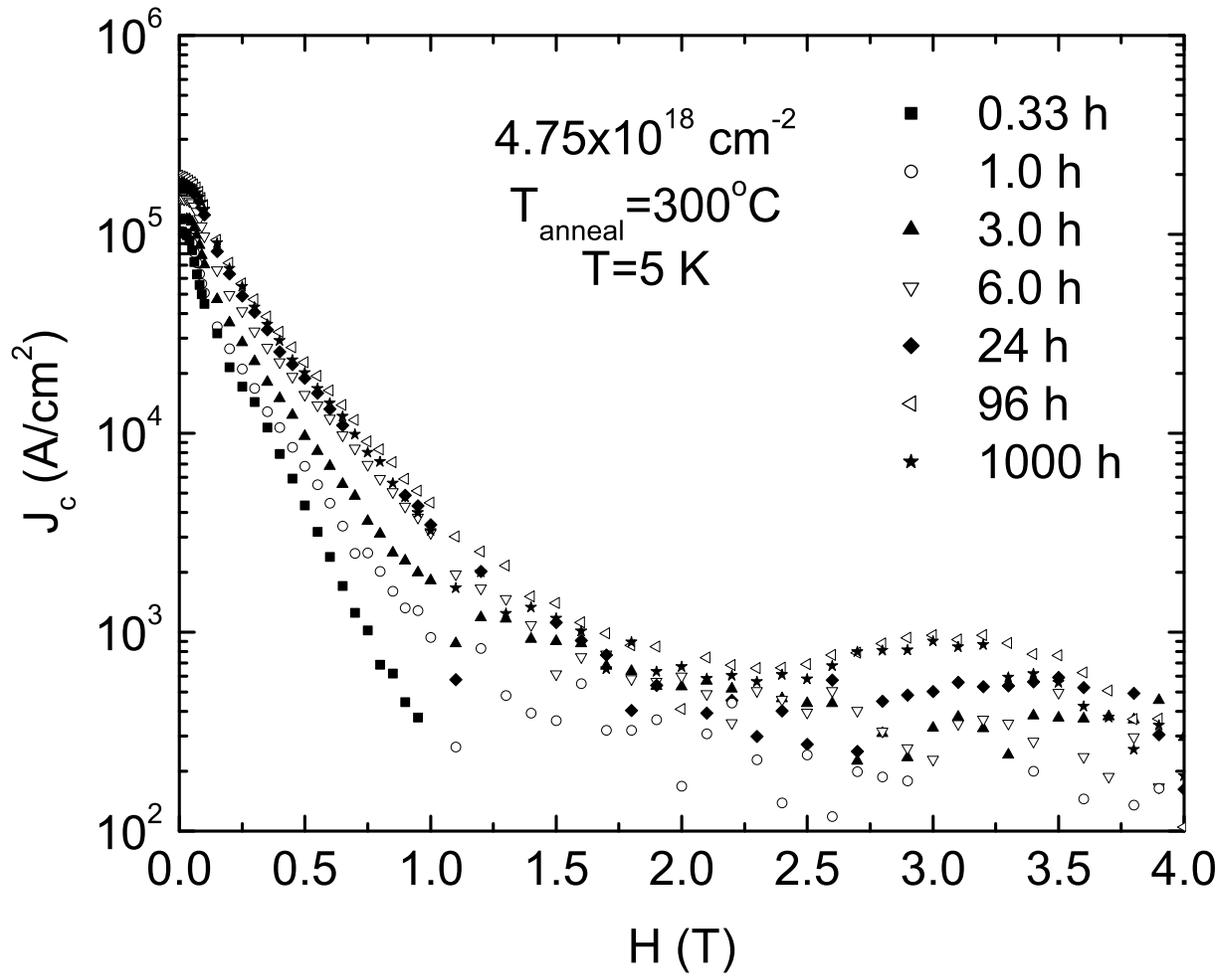}
\end{center}
\caption{$J_c$ curves (inferred from magnetization data) for the set of $4.75 \times 10^{18}$ cm$^{-2}$ fluence
samples annealed at 300$^o$C for various times.}\label{f10}
\end{figure}

\clearpage

\begin{figure}
\begin{center}
\includegraphics[angle=0,width=180mm]{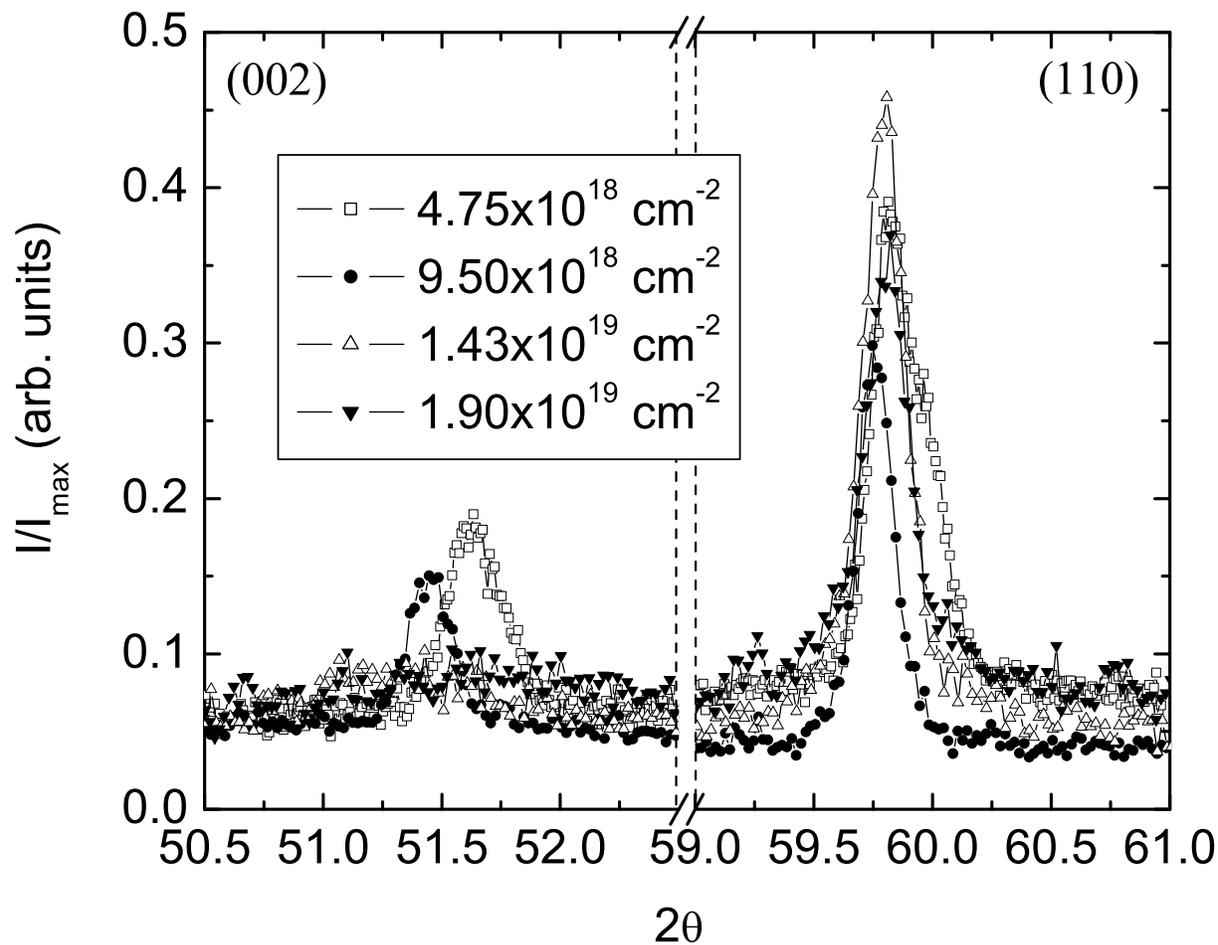}
\end{center}
\caption{(002) and (110) X-ray peaks for all four exposure levels annealed at 300$^o$C for 24 hours. The highest
two levels continue to show a substantially broadened (002) peaks indicating a degradation of long range order
along the c-direction.}\label{f11}
\end{figure}

\clearpage

\begin{figure}
\begin{center}
\includegraphics[angle=0,width=180mm]{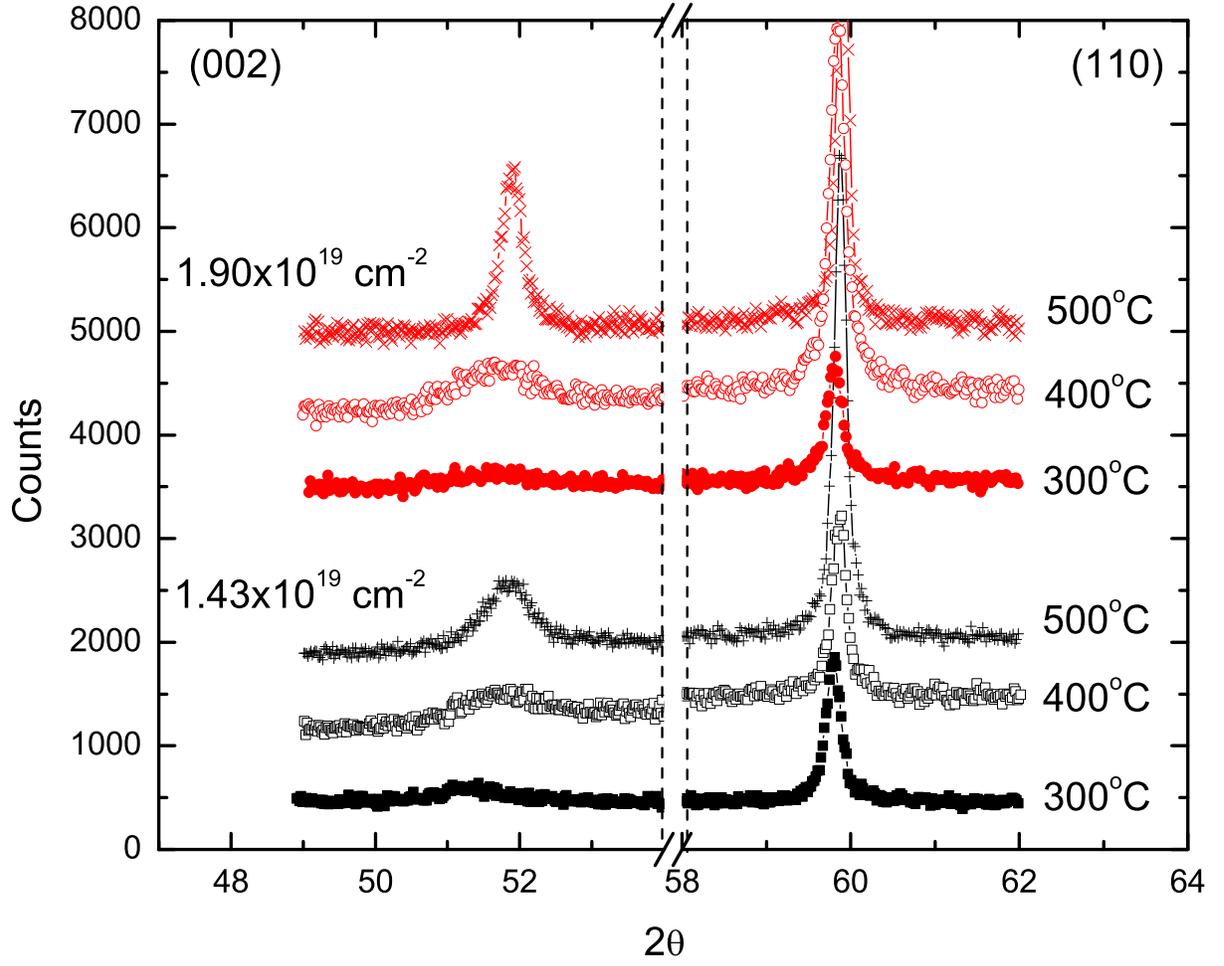}
\end{center}
\caption{(Color online) Evolution of the (002) and (110) X-ray peaks for the $1.43 \times 10^{19}$ and $1.90
\times 10^{19}$ cm$^{-2}$ exposure levels annealed for 24 hours with the annealing temperature increasing from
300$^o$C to 500$^o$C. In both cases, the (002) attains a FWHM less than 1$^o$ 2$\theta$ only after the annealing
temperature reaches 500$^o$C. Note: intermediate temperature anneals at 400$^o$C were performed for longer times
to show the persistence of the broadening for T$\leq$ 400$^o$C.}\label{f12}
\end{figure}

\clearpage

\begin{figure}
\begin{center}
\includegraphics[angle=0,width=180mm]{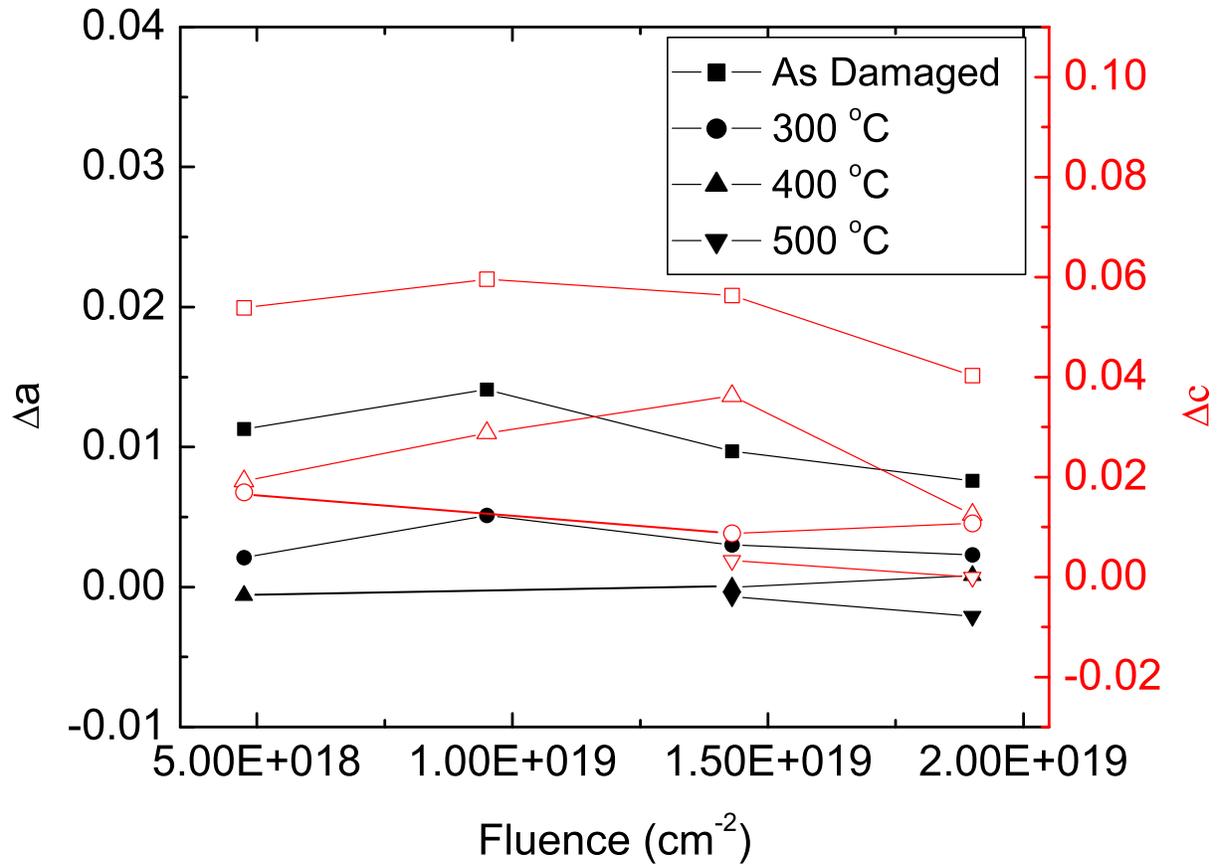}
\end{center}
\caption{(Color online) Calculated lattice parameter shifts for all four damage levels annealed at temperatures up
to 500$^o$C. Closed symbols represent $\Delta$a, and open symbols are $\Delta$c. Lines serve as guides to the
eye.}\label{f13}
\end{figure}

\clearpage

\begin{figure}
\begin{center}
\includegraphics[angle=0,width=180mm]{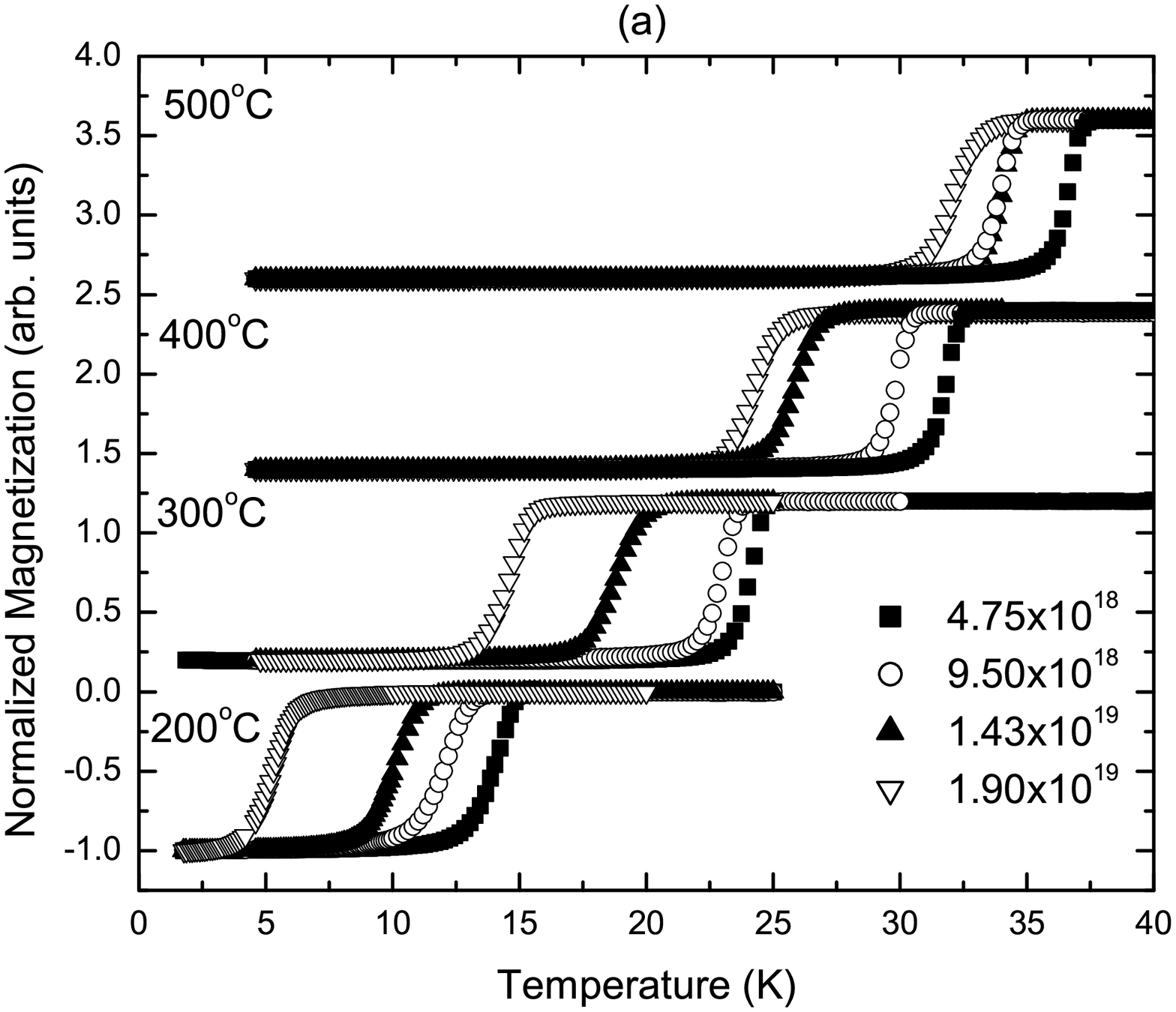}
\end{center}
\end{figure}

\clearpage

\begin{figure}
\begin{center}
\includegraphics[angle=0,width=180mm]{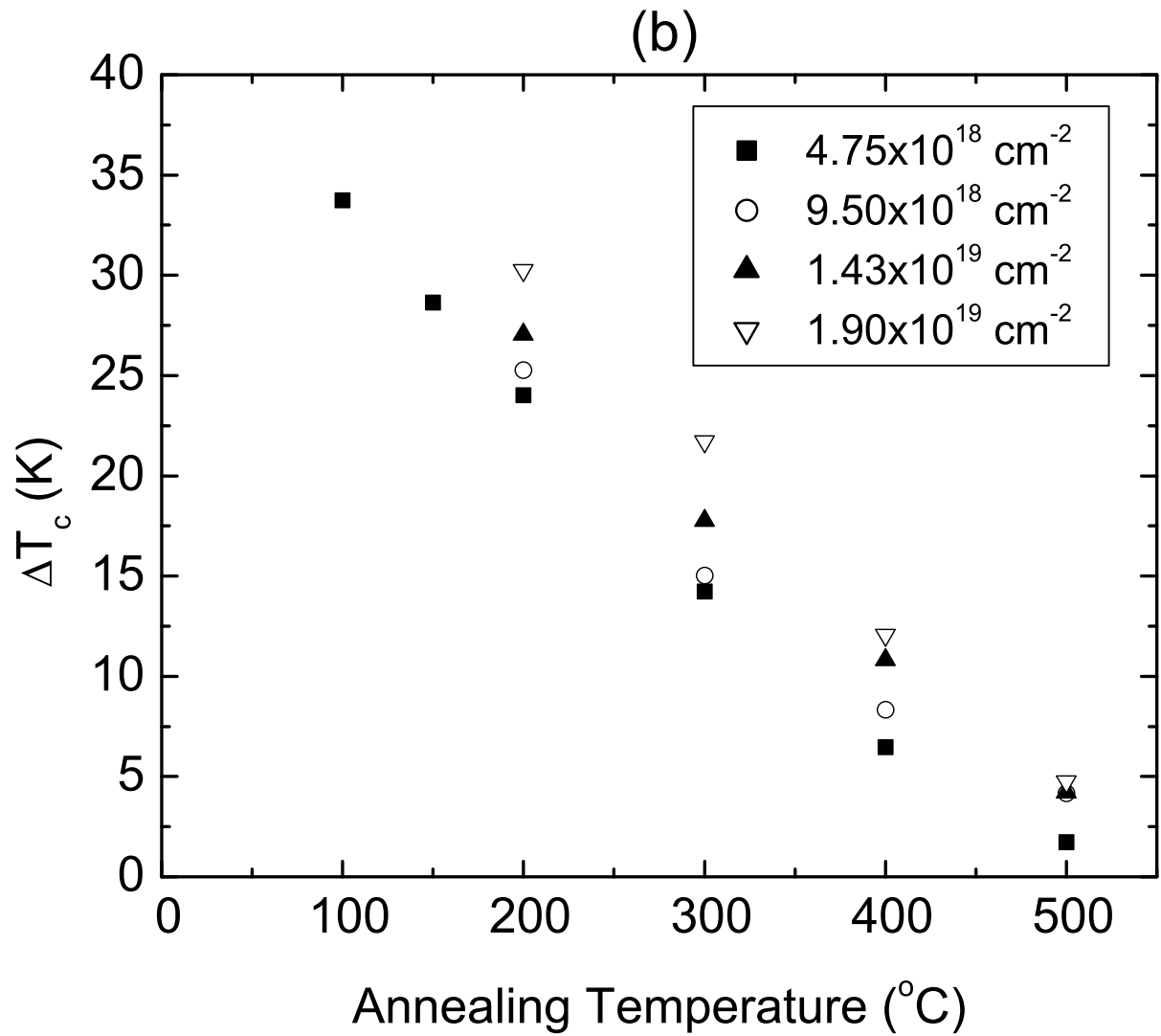}
\end{center}
\caption{(a) Normalized magnetic transitions as a function of annealing temperature and fluence level for all four
damage levels. The time for each anneal was 24 hours. The curves are normalized to a full screening value of -1.
The set of curves for each successive annealing temperature is shifted upward by 1.2 units for ease of reading.
(b) $\Delta T_c$ values, as determined by a 1\% screening criteria, as a function of annealing temperature and
fluence level. }\label{f14}
\end{figure}

\clearpage

\begin{figure}
\begin{center}
\includegraphics[angle=0,width=180mm]{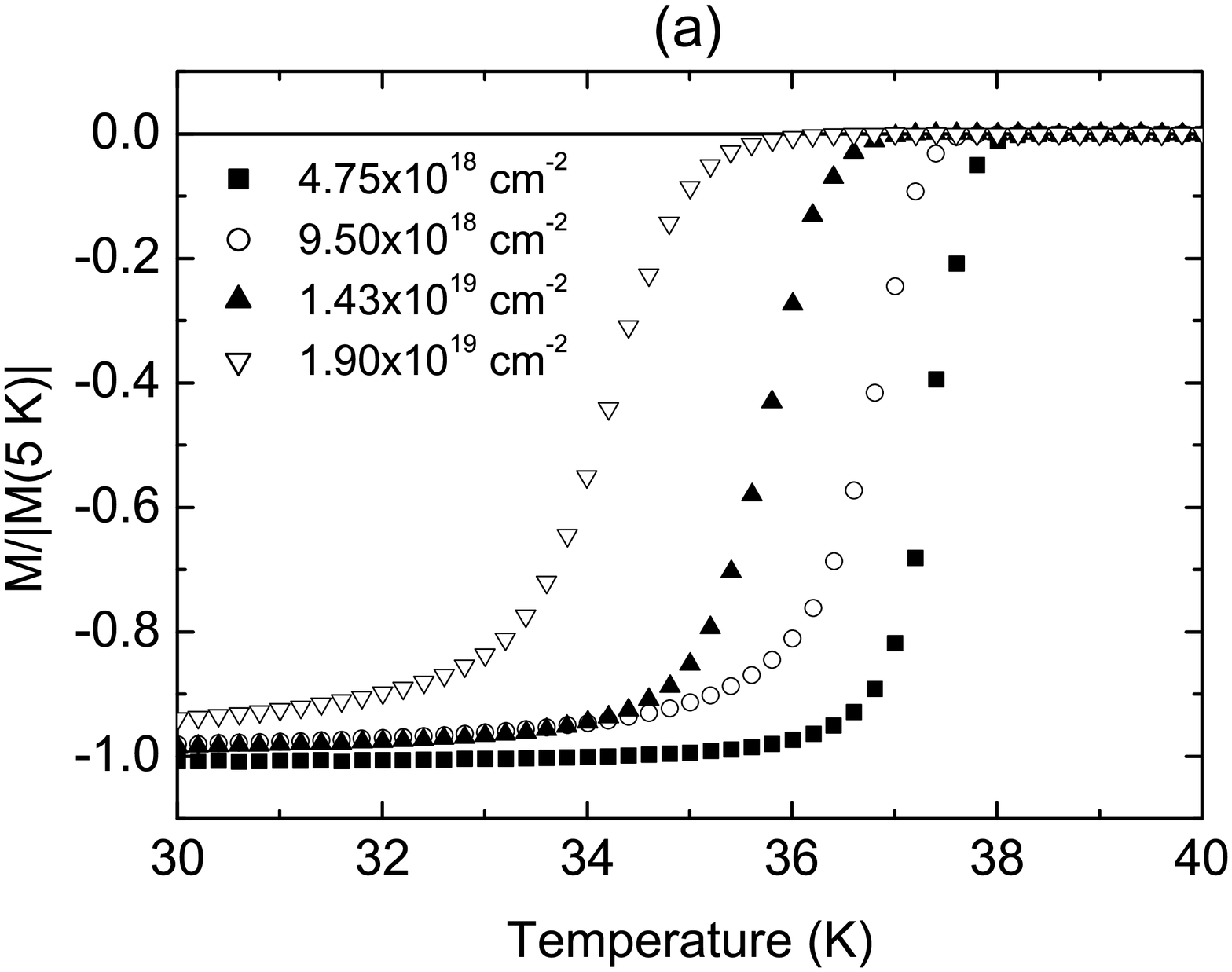}
\end{center}
\end{figure}

\clearpage

\begin{figure}
\begin{center}
\includegraphics[angle=0,width=180mm]{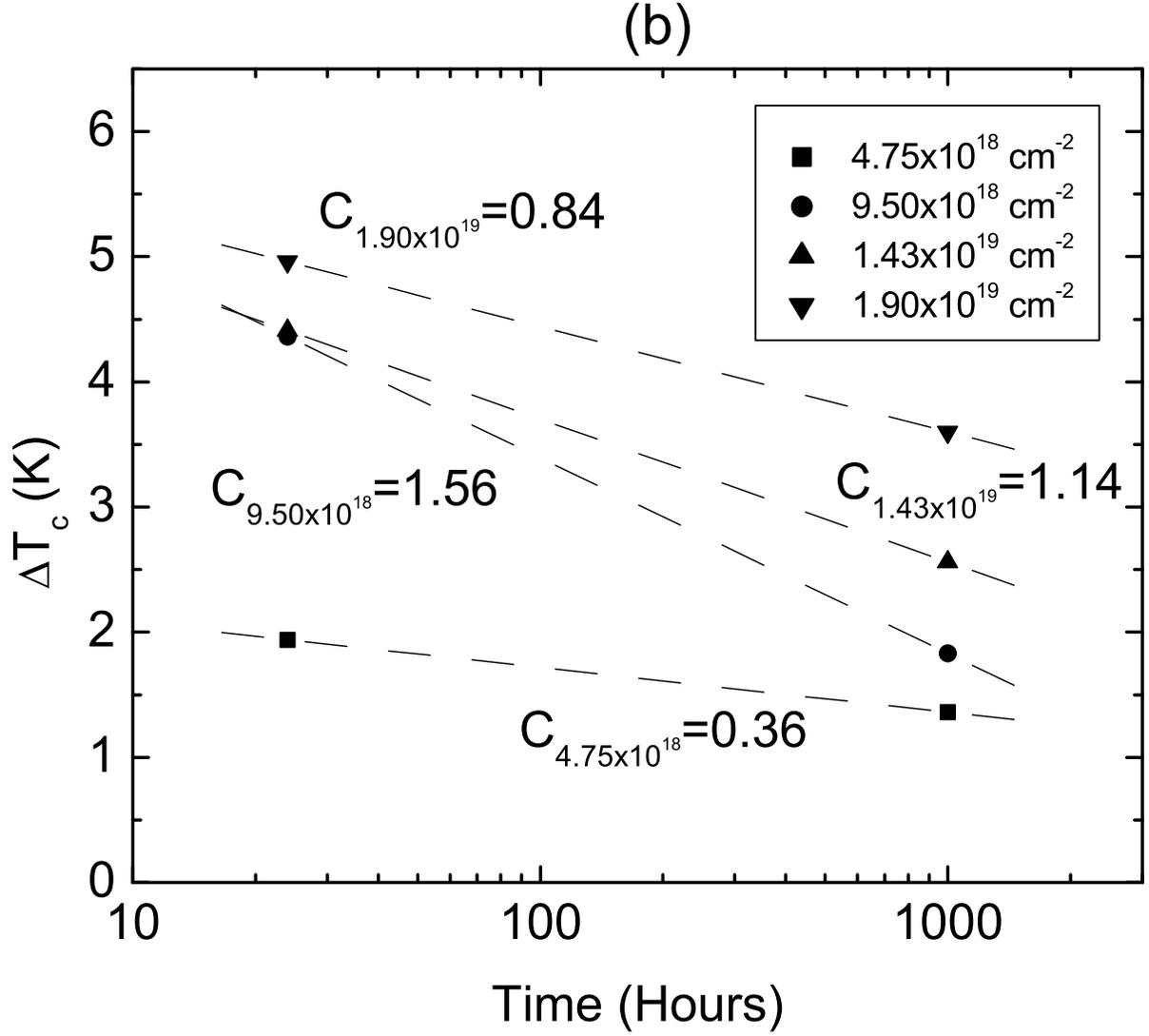}
\end{center}
\caption{(a) Normalized magnetic transitions for all four fluence levels annealed at 500$^o$C for 1000 hours. (b)
$\Delta T_c$ for samples annealed at 500$^o$C for 24 and 1000 hours. For each of the four fluence levels,
extending the annealing time results in a further recovery of $T_c$ towards that of the undamaged sample. Rate
constants, $C$, are determined by a linear fit of the semi-log plot assuming $\Delta T_c$ follows a decaying
exponential as a function of time. }\label{f15}
\end{figure}

\clearpage

\begin{figure}
\begin{center}
\includegraphics[angle=0,width=180mm]{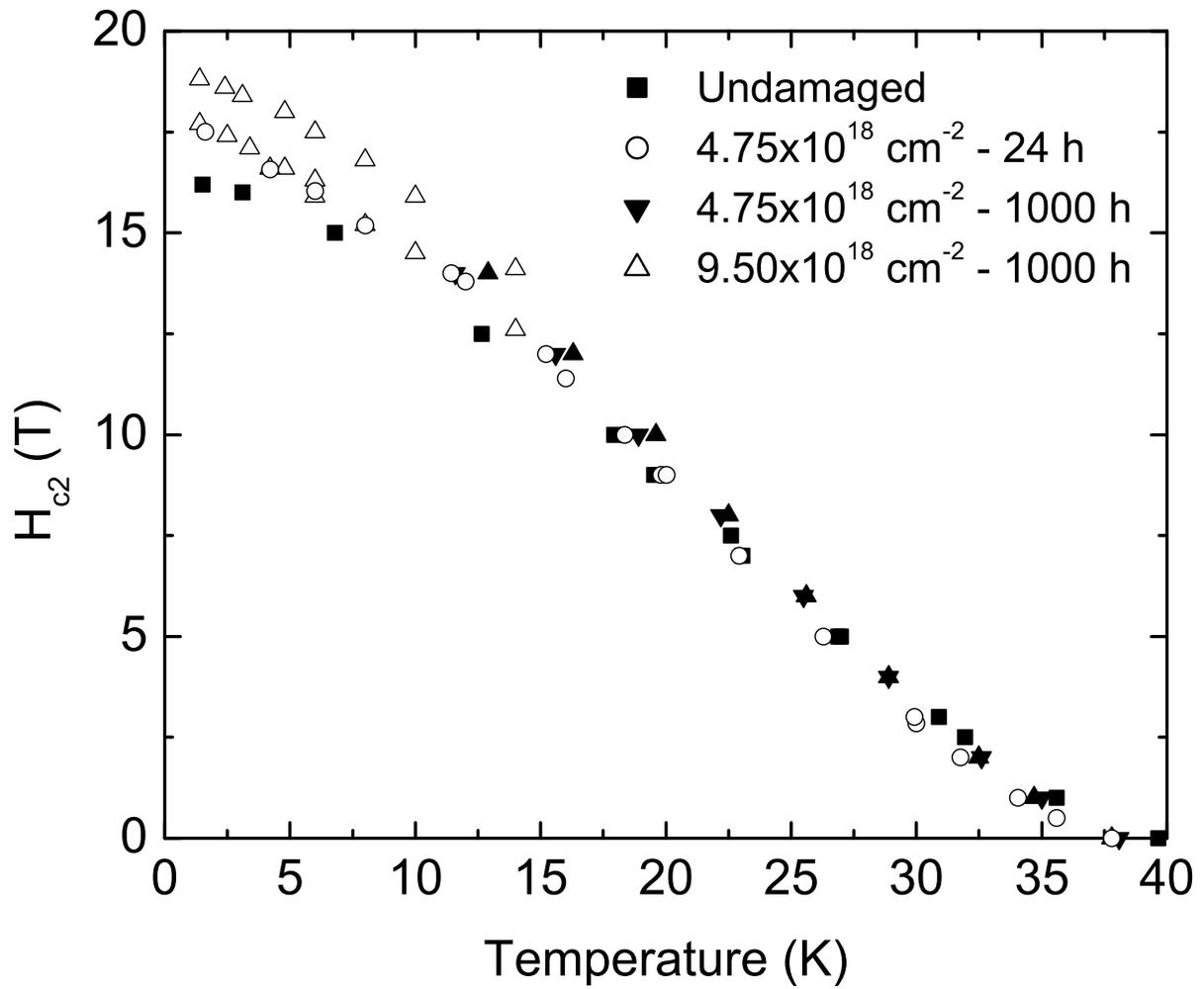}
\end{center}
\caption{$H_{c2}$ curves for the $4.75 \times 10^{18}$ cm$^{-2}$ and $9.50 \times 10^{18}$ cm$^{-2}$ fluence
levels annealed for 1000 hours at 500$^o$C. Two sets of data for the low temperature $H_{c2}$ values of the
samples exposed to a fluence of $9.50 \times 10^{18}$ cm$^{-2}$ are shown. Both damage levels show a possible
enhancement relative to the undamaged sample, but there is some spread in the data as illustrated by the two $9.50
\times 10^{18}$ cm$^{-2}$ fluence level samples shown.}\label{f16}
\end{figure}

\clearpage

\begin{figure}
\begin{center}
\includegraphics[angle=0,width=180mm]{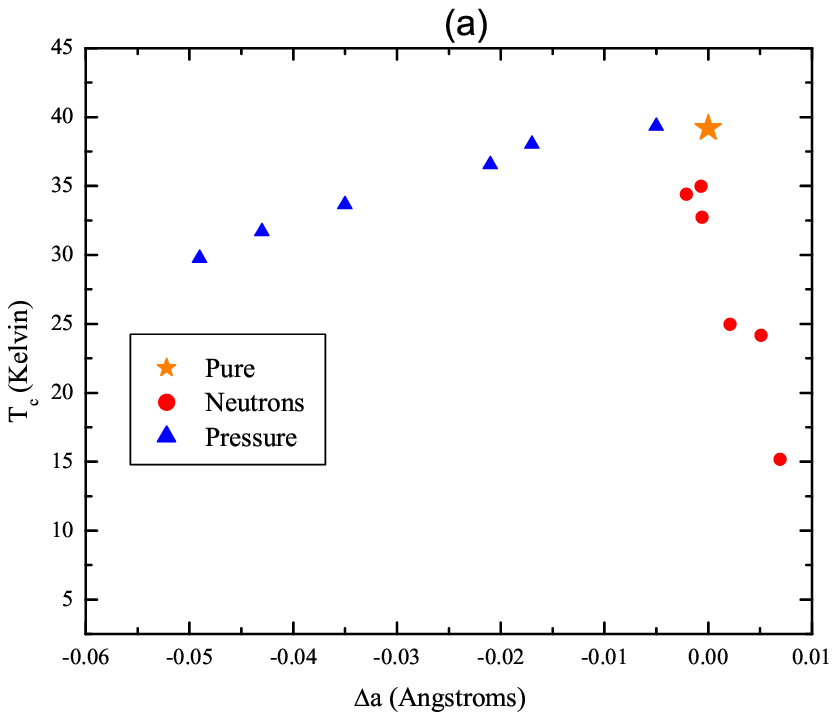}
\end{center}
\end{figure}

\clearpage

\begin{figure}
\begin{center}
\includegraphics[angle=0,width=180mm]{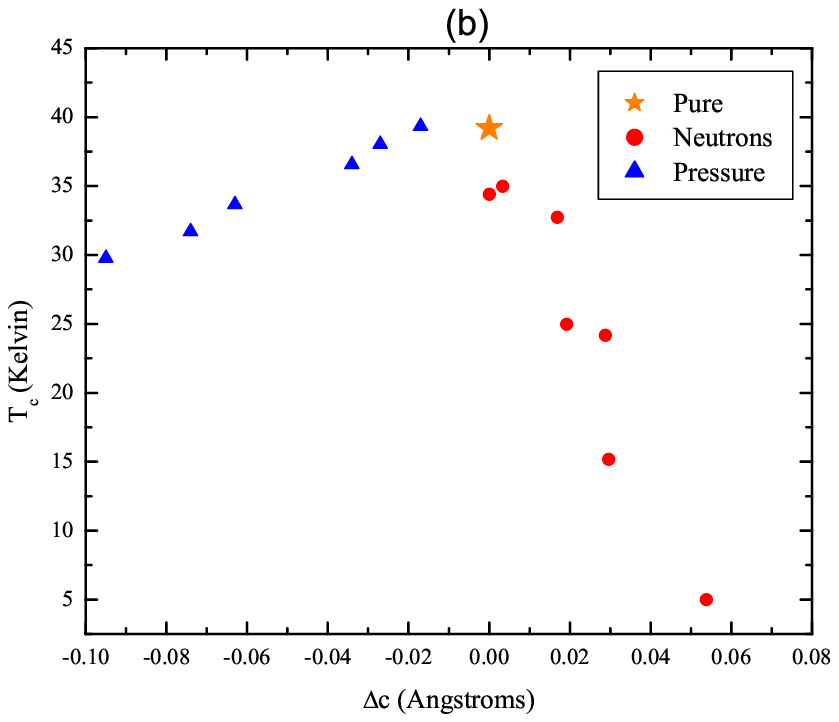}
\end{center}
\end{figure}

\clearpage

\begin{figure}
\begin{center}
\includegraphics[angle=0,width=180mm]{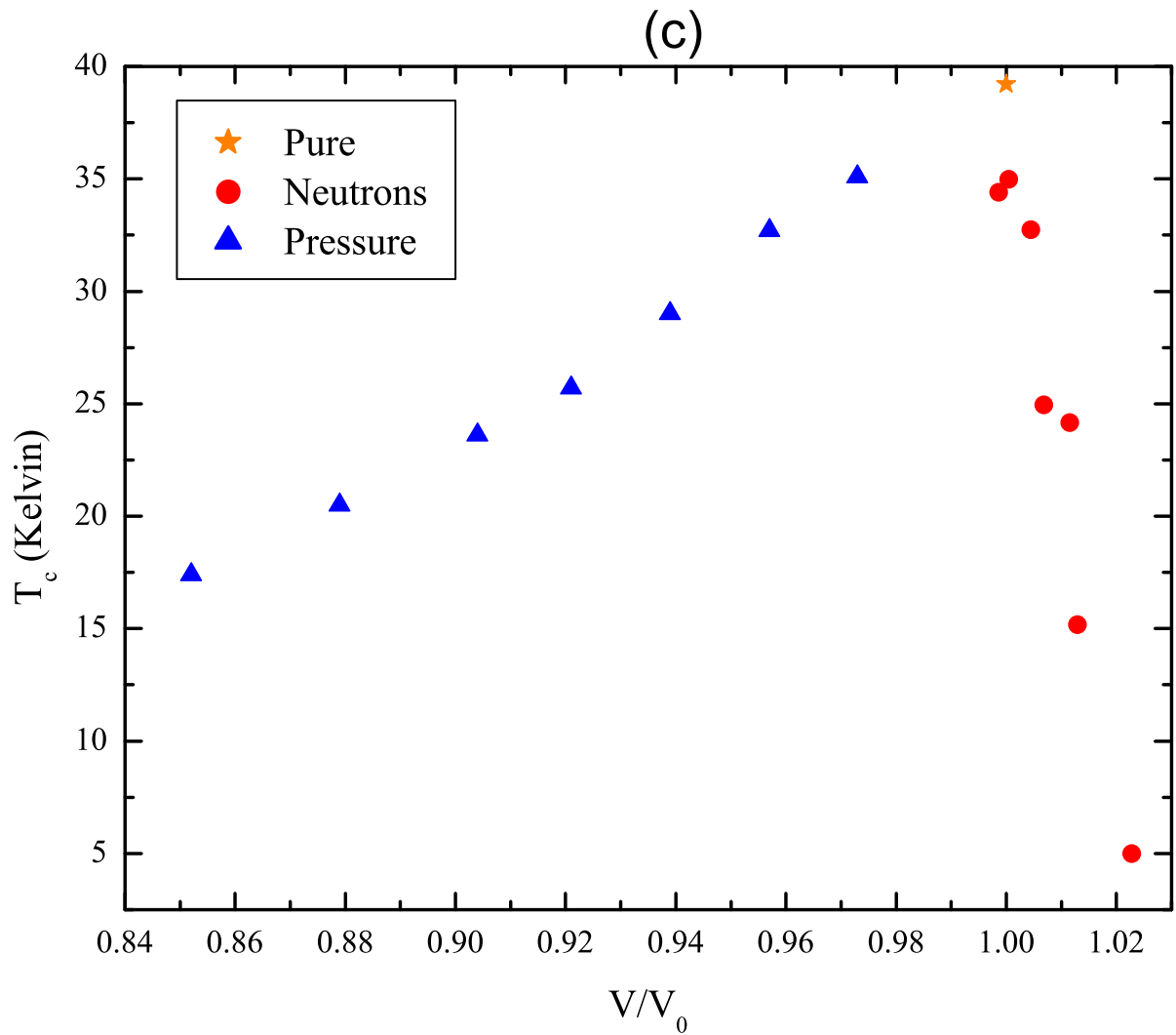}
\end{center}
\caption{(Color online) Comparison of the different development of $T_c$ with (a) $\Delta$a, (b) $\Delta$c, and
(c) unit cell volume for neutron irradiated MgB$_2$ and MgB$_2$ under external pressure. Pressure data are
recreated from reference \cite{34}. }\label{f17}
\end{figure}

\clearpage

\begin{figure}
\begin{center}
\includegraphics[angle=0,width=180mm]{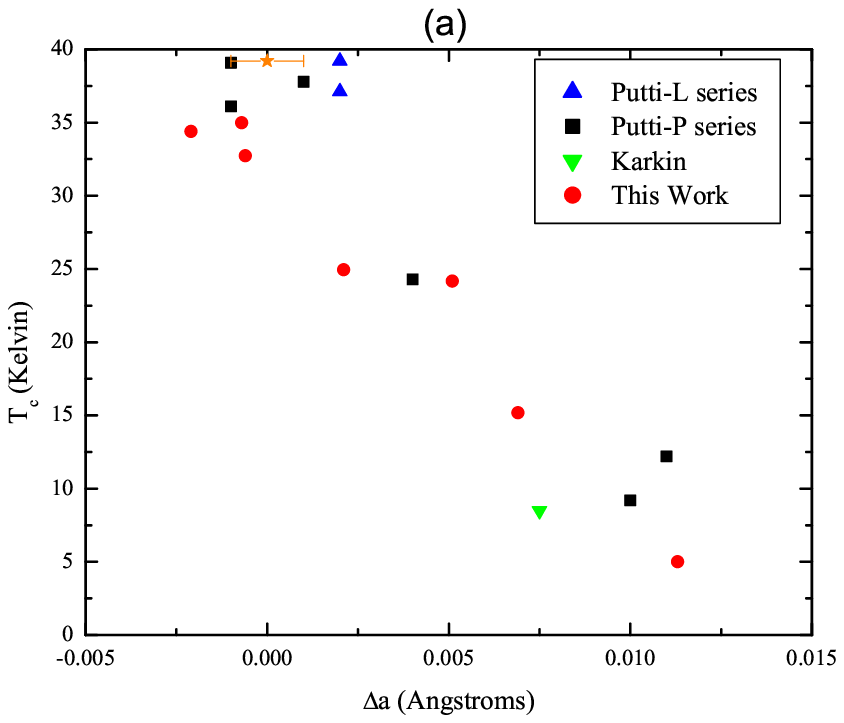}
\end{center}
\end{figure}

\clearpage

\begin{figure}
\begin{center}
\includegraphics[angle=0,width=180mm]{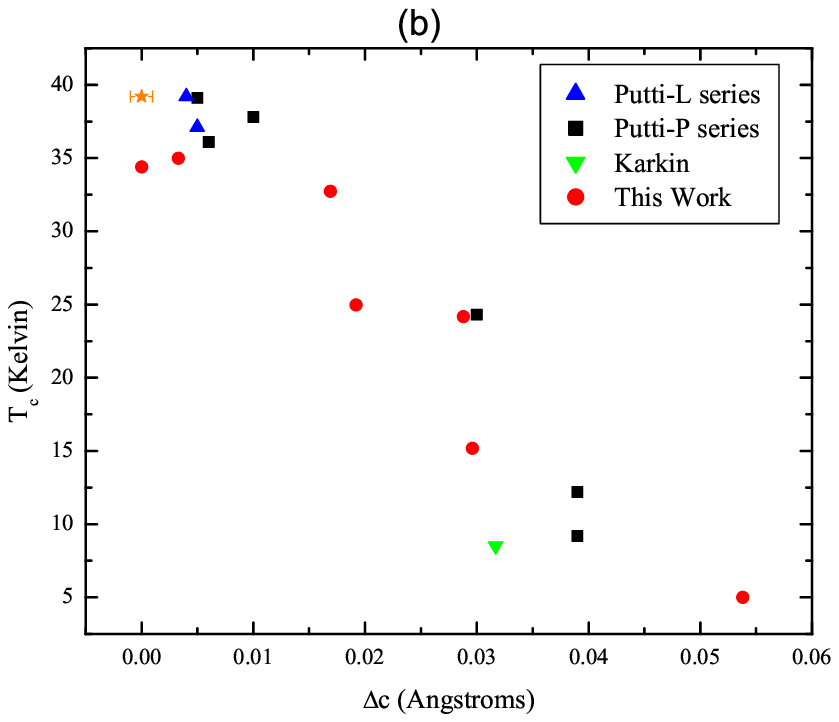}
\end{center}
\end{figure}

\clearpage

\begin{figure}
\begin{center}
\includegraphics[angle=0,width=180mm]{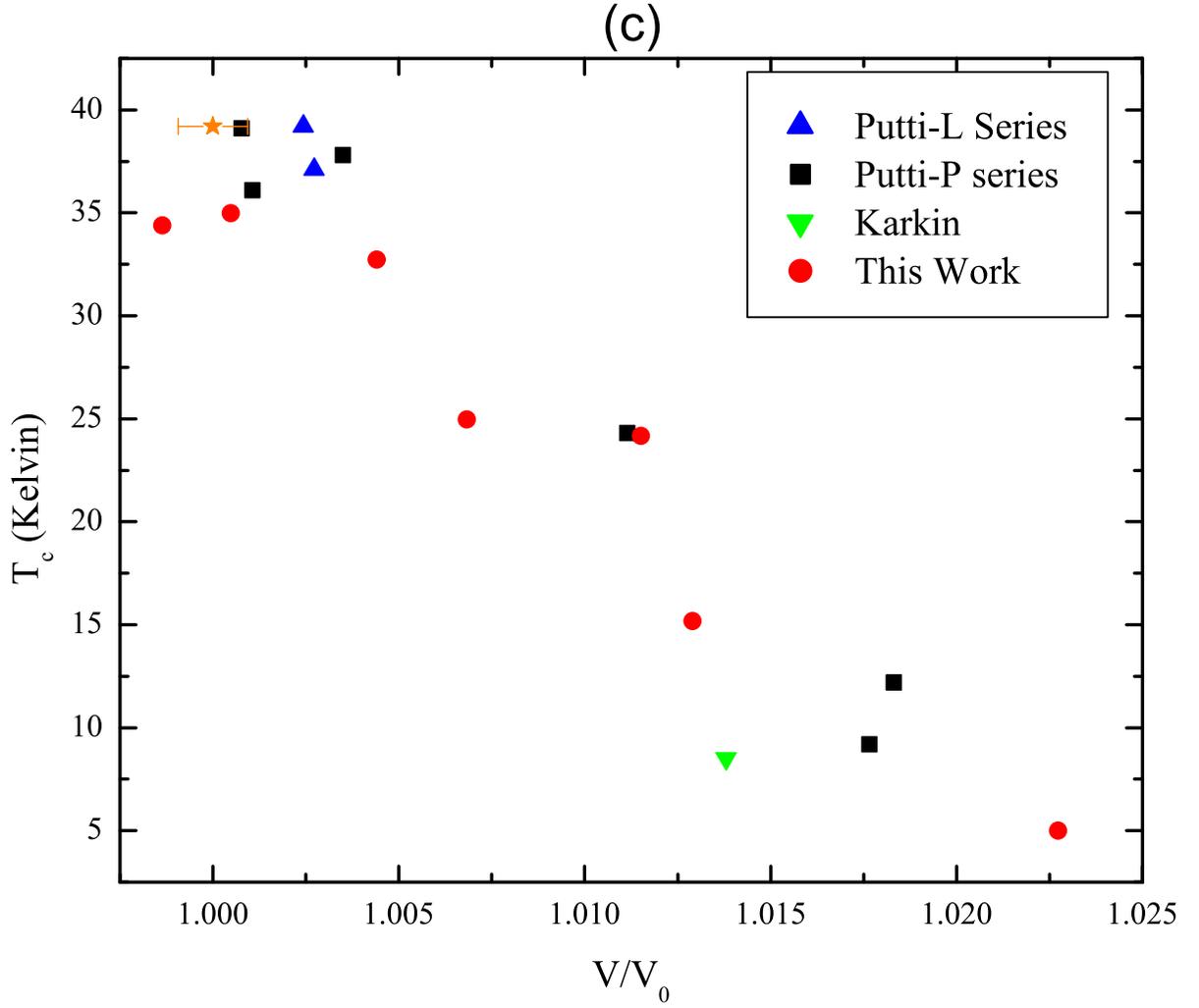}
\end{center}
\caption{(Color online) Comparison of the different development of $T_c$ with (a) $\Delta$a, (b) $\Delta$c, and
(c) unit cell volume for neutron irradiation from different groups. Data includes results from references
\cite{14,20}. In parts (a) and (b) the error bars on the pure sample represent typical experimental error in
determining lattice parameters. In the part (c), the error bars represent the propagation of the error in the
lattice parameters to the calculated $V/V_0$ value. All of our data in these plots should be considered to have
comparable uncertainty.}\label{f18}
\end{figure}

\clearpage

\begin{figure}
\begin{center}
\includegraphics[angle=0,width=180mm]{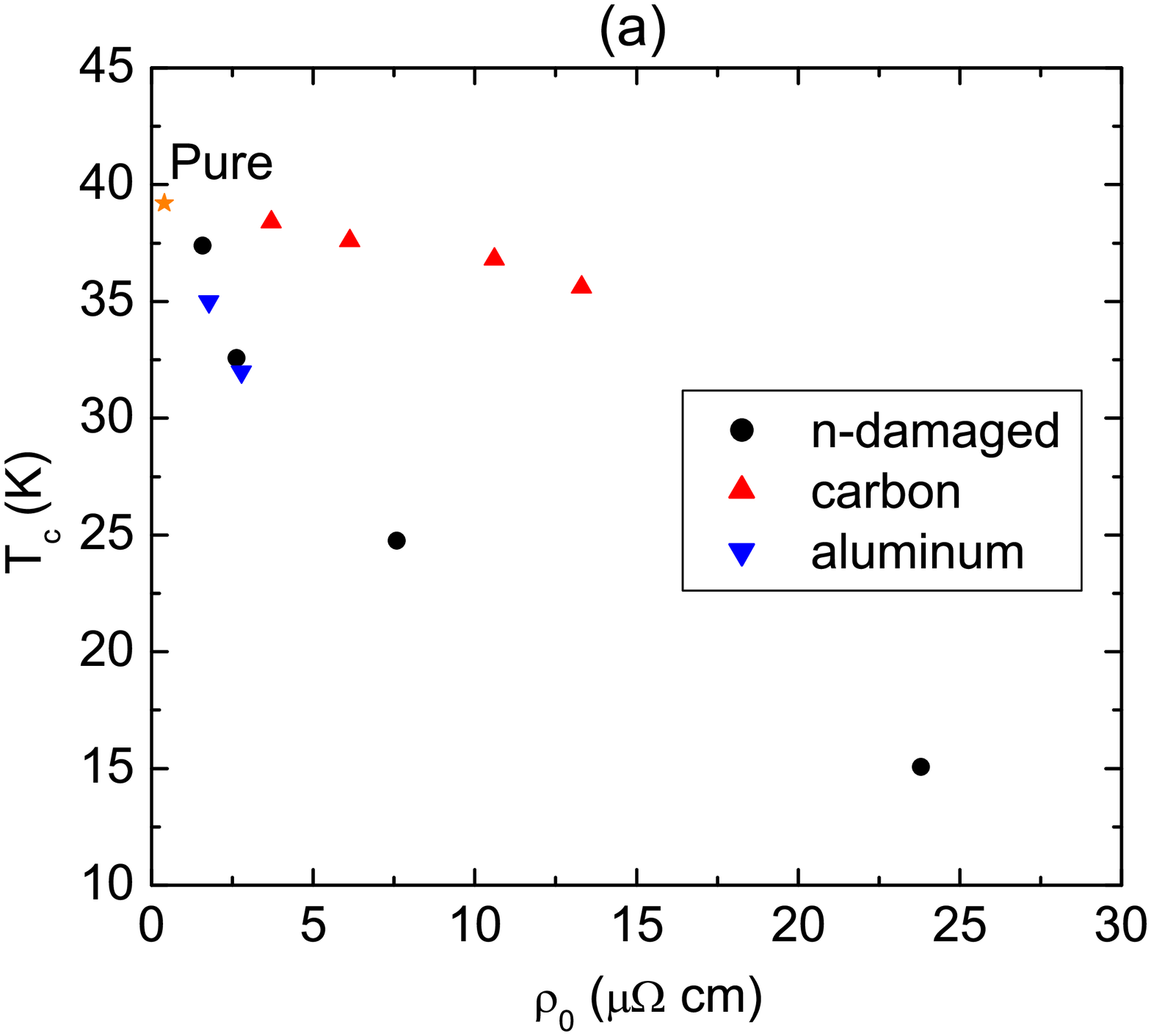}
\end{center}
\end{figure}

\clearpage

\begin{figure}
\begin{center}
\includegraphics[angle=0,width=180mm]{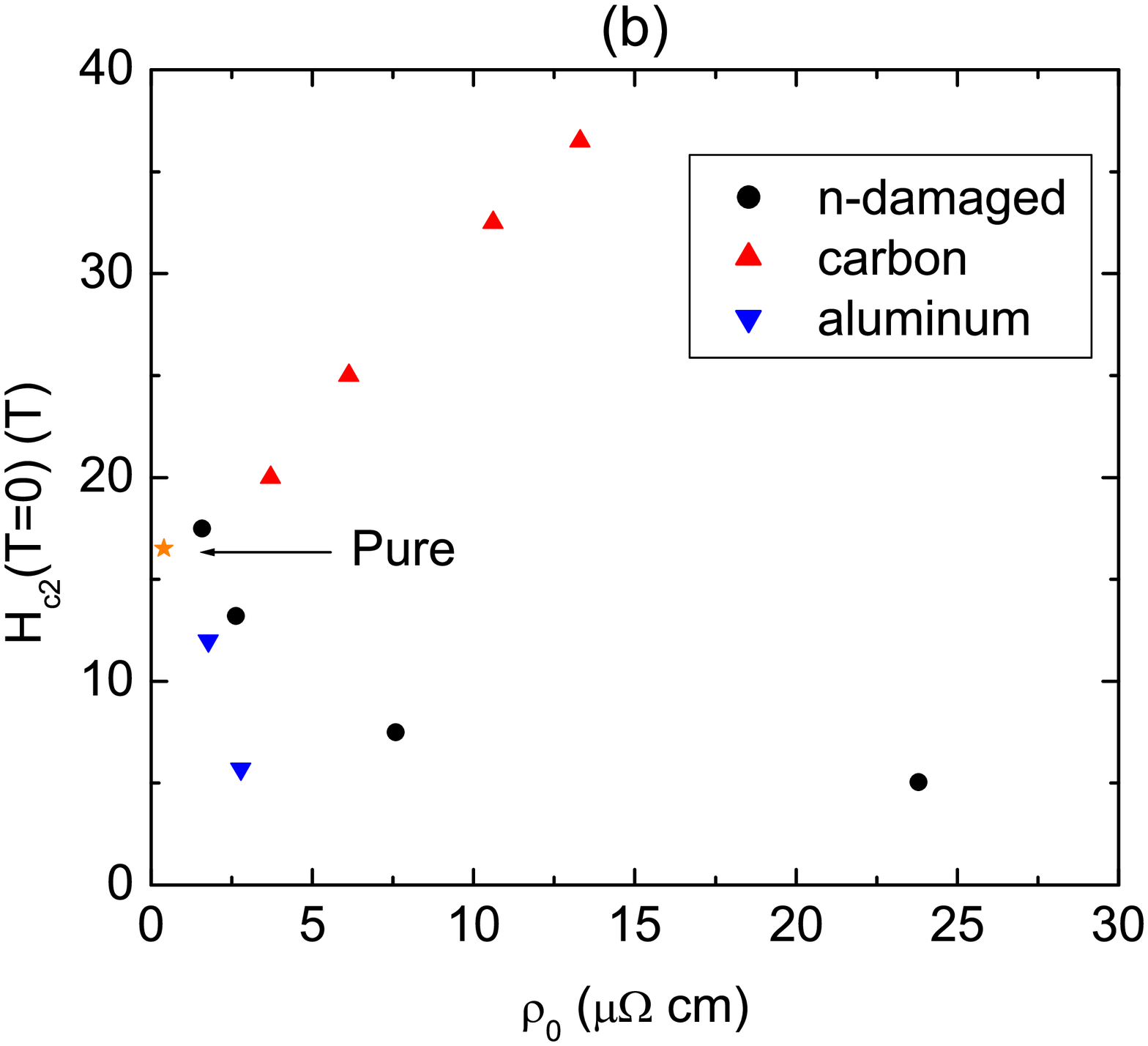}
\end{center}
\end{figure}

\clearpage

\begin{figure}
\begin{center}
\includegraphics[angle=0,width=180mm]{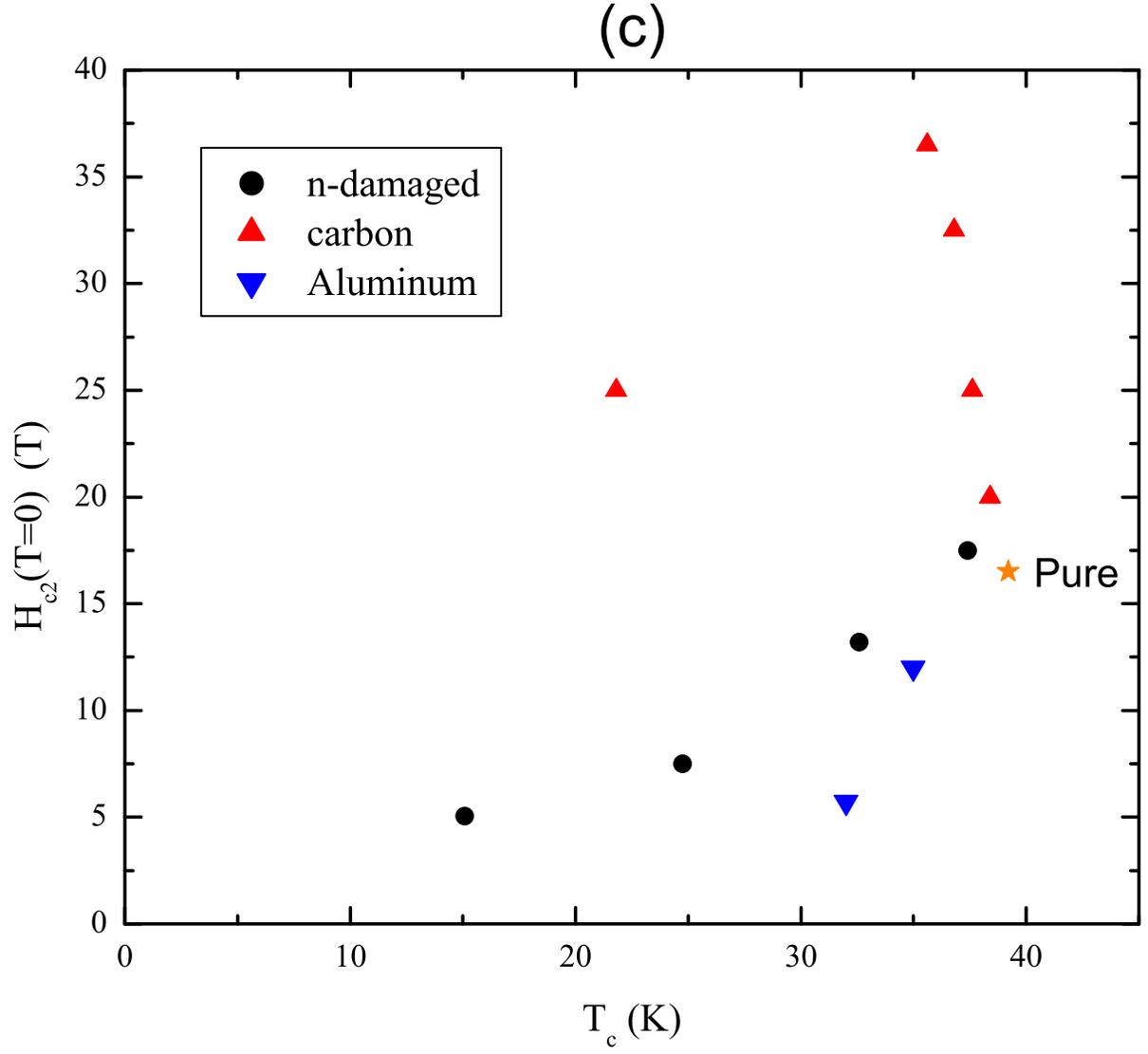}
\end{center}
\caption{(Color online) Interdependencies of $T_c$, $H_{c2}$(T=0), and $\rho$$_0$ for carbon doped \cite{187,188},
aluminum doped single crystals \cite{9}, and neutron irradiated samples. (a) $T_c$ vs. $\rho$$_0$, (b)
$H_{c2}(T=0)$ vs. $\rho$$_0$, and (c) $H_{c2}(T=0)$ vs. $T_c$.}\label{f19}
\end{figure}

\clearpage

\begin{table}
 \begin{center}

\begin{tabular}{|c|c|c|} \hline
Exposure Time (Hours) & Fluence (cm$^{-2}$) & Atomic \% Li \\
\hline
24 & $4.75 \times 10^{18}$ & 0.37\\
48 & $9.50 \times 10^{18}$ & 0.74\\
72 & $1.43 \times 10^{19}$ & 1.11\\
96 & $1.90 \times 10^{19}$ & 1.48\\
\hline
\end{tabular}
 \caption{Estimated fluences for each of the four exposure times
and calculated atomic percentage of boron transmutated to lithium.}\label{t1}
\end{center}
\end{table}


\begin{thebibliography}{00}
\bibitem {1} J. Nagamatsu, N. Nakagawa, T. Muranaka, Y. Zenitani, J.
Akimitsu, Nature (London) 2001 (410) 6824.
\bibitem {2} S.L. Bud'ko, G. Lapertot, C. Petrovic, C.E. Cunngingham, N. Anderson,
and P.C. Canfield, Phys. Rev. Lett. 86 (2001) 1877-1880.
\bibitem {3} J. Kortus, I. I. Mazin, K. D. Belashchenko, V. P. Antropov, and L. L.
Boyer, Phys. Rev. Lett. 86 (2001) 4656.
\bibitem {4} R.A. Fisher, G. Li, J.C. Lashley, F. Bouquet, N.E. Phillips, D.G. Hinks,
J.D. Jorgensen, and G.W. Crabtree, Physica C 385 (2003) 49.
\bibitem {5} Physica C 385 (2003), edited by G. Crabtree, W. Kwok, S.L. Bud'ko, and P.C. Canfield.
\bibitem {6} R.J. Cava, H.W. Zandergen, K. Inumaru, Physica C 385 (2003) 8.
\bibitem {7} M. Angst, S.L. Bud'ko, R.H.T. Wilke, and P.C. Canfield, Phys. Rev. B 71 (2005) 144512.

\bibitem {8} B. Kang, H.J. Kim, H.S. Lee, S.I. Lee, and T. Dahm,
cond-mat/0409496. 

\bibitem {9} J. Karpinski, N.D. Zhigadlo, G. Schuck. S.M. Kazakov, B. Batlogg, K. Rogacki,
R. Puzniak, J. Jun, E. Muller, P. Wagli, R. Gonnelli, D. Daghero, G.A. Ummarino, V.A. Stepanov, and P.N. Lebedev,
Phys. Rev. B 71 (2005) 174506.
\bibitem {10} M. Putti, C. Ferdeghini, M. Monni, I. Pallecchi, C.
Tarantini, P. Manfrinetti, A. Palenzona, D. Daghero, R.S. Gonnelli, and V.A. Stepanov, Phys. Rev. B 71 (2005)
144505.
\bibitem {11} R.H.T. Wilke, S.L. Bud'ko, P.C. Canfield, D.K.
Finnemore, Raymond J. Suplinskas, and S.T. Hannahs, Phys. Rev. Lett. 91 (2004) 217003.

\bibitem {12} R. Puzniak, M. Angst, A. Szewczyk, J. Jun, S.M.
Kazakov, and J. Karpinski, cond-mat/0404579. 

\bibitem {13} A. Gurevich, Phys. Rev. B. 67 (2003) 184515.
\bibitem {24} A.C. Damask and G.J. Dienes, Point Defects in Metals,
Gordon and Breach, Science Publishers, Inc., New York, 1963.
\bibitem {14} A.E. Karkin, V.I. Voronin, T.V. Dyachkova,
A.P. Tyutyunnik, V.G. Zubkov, Yu. G. Zainulin, and B.N. Goshchitskii, JETP Letters 73 (2001) 570.
\bibitem {15} M. Eisterer, M. Zehetmayer, S. T\"{o}nies, H. W. Weber,
M. Kambara, N. Hari Babu, D.A. Cardwell, and L.R. Greenwood, Superconductor Science and Technology 15 (2002)
L9-L12.
\bibitem {16} E. Babic, D. Miljanic, K. Zadro, I. Kusevic, Z. Marohnic,
D. Drobac, X.L. Wang, and S.X. Dou, Fizika A 10 (2001) 87-94.
\bibitem {17} U.P. Trociewitz, P.V.P.S.S. Sastry, A. Wyda, K. Crockett, and
J. Schwartz, IEEE Transactions on Applied Superconductivity 13 (2003) 3320-3323.
\bibitem {99} Kenneth S. Krane, Introduction to Nuclear Physics,
John Wiley \& Sons, Inc., New York, 1988.
\bibitem {18} Y. Wang, F. Bouquet, I. Sheikin, P. Toulemonde, B. Revaz,
M. Eisterer, H.W. Weber, J. Hinderer, and A. Junod, J. Phys.: Condens. Matter 15 (2003) 883-893.
\bibitem {19} M. Zehetmayer, M. Eisterer, J. Jun, S.M. Kazakov, J. Karpinski,
B. Birajdar, O. Eibl, and H.W. Weber, Phys. Rev. B 69 (2004) 054510.
\bibitem {20} M.Putti, V. Braccini, C. Ferdeghini, F. Gatti, P. Manfrinetti,
D. Marre, A. Palenzona, I. Pallecchi, C. Tarantini, I. Sheikin, H.U. Aebersold, and E. Lehmann, App. Phys. Lett.
86 (2005) 112503.
\bibitem {21} M. Ortolani, D. Di Castro, P. Postorino, I. Pallecchi, M. Monni,
M. Putti, and P. Dore, Phys. Rev. B 71 (2005) 172508.
\bibitem {22} P.C. Canfield, D.K. Finnemore, S.L. Bud'ko, J.E. Ostenson, G. Lapertot,
C.E. Cunningham, and C. Petrovic, Phys. Rev. Lett. 86 (2001) 2423.

\bibitem {187} R.H.T. Wilke,  S.L. Bud'ko, P.C. Canfield, D.K.
Finnemore, Raymond J. Suplinskas, and S.T. Hannahs, Physica C 424 (2005) 1.

\bibitem {730} R.A. Ribeiro, S.L. Bud'ko, C. Petrovic, and P.C. Canfield, Physica C 382 (2002) 194.
\bibitem {991} R.H.T. Wilke, S.L. Bud'ko, P.C. Canfield, D.K. Finnemore, Raymond J. Suplinskas,
J. Farmer, and S.T. Hannahs, cond-mat/0507275

\bibitem {476} R. Gandikota, R.K. Singh, J. Kim, B. Wilkens, N.
Newman, J.M. Rowell, A.V. Pogrebnyakov, X.X. Xi, J.M. Redwing, S.Y. Xu, and Q. Li, App. Phys. Lett. 86 (2005)
012508.
\bibitem {1215} I.I. Mazin, O.K. Andersen, O. Jepsen, O.V. Dolgov,
J. Kortus, A.A. Golubov, A.B. Kuz'menko, and D. van der Marel, Phys. Rev. Lett. 89 (2002) 107002.
\bibitem {1216} J.M. Rowell, Supercond. Sci. Tech. 16 (2003) R17.
\bibitem {1217} M. Eisterer, Phys. Stat. Sol. C 2 (2005) 1606.


\bibitem {25} N.R. Werthamer, E. Helfand, and P.C. Hohenberg, Phys. Rev. 147 (1966) 295.
\bibitem {1218} M. Putti, M. Affronte, C. Ferdeghini, C.
Tarantini, and E. Lehmann, cond-mat/050852

\bibitem {23} C.P. Bean, Phys. Rev. Lett 8 (1962) 250.
\bibitem {1219} M. Eisterer, M. Zehetmayer, and H.W. Weber, Phys.
Rev. Lett. 90 (2003) 247002.
\bibitem {1230} I. Pallecchi, J.C. Tarantini, H.U. Aebersold, V.
Braccini, C. Fanciulli, C. Ferdeghini, F. Gatti, E. Lehmann, P. Manfrinetti, D. Marre', A. Palenzona, A.S. Siri,
M. Vignolo, and M. Putti, Phys. Rev. B 71 (2005) 212507.

\bibitem {27} S.X. Dou, S. Soltanian, J. Horvat, X.L. Wang, S.H.
Zhou, M. Ionescu, H.K. Liu, P. Munroe, and M. Tomsic, APL 81 (2002) 3419.
\bibitem {106} M.W. Thompson, Defects and Radiation Damage in
Metals, Cambridge University Press, London, 1969.
\bibitem {69} D.K. Finnemore, J.E. Ostensen, S.L. Bud'ko, G.
Lapertot, and P.C. Canfield, Phys. Rev. Lett. 86 (2001) 2420.
\bibitem {igor} A.A. Golubov and I.I. Mazin, Phys. Rev. B 55 (1997) 15146.
\bibitem {823} M.R. Eskildsen, M. Kugler, S. Tanaka, J. Jun, S.M. Kazakov, J. Karpinksi,
and $\O$. Fischer, Phys. Rev. Lett. 89 (2002) 187003.
\bibitem {34} A.F. Goncharov and V.V. Struzhkin, Physica C 385
(2003) 117.
\bibitem {35} X.H. Zeng, A.V. Pogrebnyakov, M.H. Zhu, J.E. Jones,
X.X. Xi, S.Y. Xu, E. Wertz, Qi Li, J.M. Redwing, J. Lettieri, V. Vaithyanathan, D.G. Schlom, Zi-Kui Liu, O.
Trithaveesak, and J. Schubert, Appl. Phys. Lett. 82 (2003) 2097.
\bibitem {36} S. Deemyad, T. Tomita, J.J. Hamlin, B.R. Beckett,
J.S. Schilling, J.D. Jorgensen, S. Lee, and S. Tajima, Physica C 385 (2003) 105.
\bibitem {84} V.P. Antropov, K.D. Belashchenko, M. van Shilfgaarde,
and S.N. Rashkeev, Studies of High Temperature Superconductors 38 (2002) 91.
\bibitem {29} A.P. Gerashenko, K.N. Mikhalev, S.V. Verkhovskii,
A.E. Karkin, and B.N. Goshchitskii, Phys. Rev. B 65 (2002) 132506.
\bibitem {30} S. Serventi, G. Allodi, C. Bucci, R. De Renzi, G.
Guidi, E. Pavarini, P. Manfrinetti, and A. Palenzona, Supercond. Sci. Technol. 16 (2003) 152-155.
\bibitem {31} S.H. Baek, B.J. Suh, E. Pavarini, R.G. Barnes, S.L.
Bud'ko, and P.C. Canfield, Phys. Rev. B 66 (2002) 104510.

\bibitem {477} R. Gandikota, R.K. Singh, J. Kim, B. Wilkens, N.
Newman, J.M. Rowell, A.V. Pogrebnyakov, X.X. Xi, J.M. Redwing, S.Y. Xu, Q. Li, and B. Moeckly, App. Phys. Lett. 87
(2005) 072507.
\bibitem {1226} A.V. Sologubenko, N.D. Zhigadlo, S.M. Kazakov, J.
Karpinski, and H.R. Ott, Phys. Rev. B 71 (2005) 020501.
\bibitem {188} Z. Ho\v{l}anov\'{a}, J. Ka\v{c}mar\v{c}\'{i}k,
P. Szab\'{o}, P. Samuely, I. Sheikin, R. A. Ribeiro, S. L. Bud'ko and P. C. Canfield, Physica C 404 (2004) 195.



\end{thebibliography}
\end{document}